\documentclass[12pt,preprint]{aastex}

\newcommand{\Msolar}{\mbox{$M_{\odot}\,$}}
\newcommand{\Lsolar}{\mbox{$L_{\odot}\,$}}

\newcommand{\arcsecs}{\mbox{$^{\prime\prime}$}}

\newcommand{\parcsec}{\mbox{$\stackrel{\prime\prime}{\textstyle .}$}}

\def\gs{\mathrel{\raise0.35ex\hbox{$\scriptstyle >$}\kern-0.6em \lower0.40ex\hbox{{$\scriptstyle \sim$}}}}
\def\ls{\mathrel{\raise0.35ex\hbox{$\scriptstyle <$}\kern-0.6em \lower0.40ex\hbox{{$\scriptstyle \sim$}}}}

\shorttitle{Molecular gas properties in Arp\,220 and NGC\,6240}
\shortauthors{Greve et al.}

\begin{document}


\title{Molecular gas in extreme star-forming environments: the starbursts Arp\,220 and NGC\,6240 as case studies}


\author{T.\ R.\ Greve\altaffilmark{1,2}}
\affil{Max-Planck Institut f\"ur Astronomie, K\"onigstuhl 17, 69117 Heidelberg, Germany}
\affil{California Institute of Technology, Pasadena, CA 91125, USA}
\email{tgreve@mpia-hd.mpg.de}
\author{P.\ P.\ Papadopoulos\altaffilmark{3}}
\affil{Argelander Institute for Astronomy, University of Bonn, Auf dem H\"ugel 71, 53121 Bonn, Germany}
\email{padeli@astro.uni-bonn.de}
\author{Y.\ Gao\altaffilmark{4}}
\affil{Purple Mountain Observatory, Chinese Academy of Sciences, Nanjing 210008, People's Republic of China}
\email{yugao@pmo.ac.cn}
\author{S.\ J.\ E.\ Radford\altaffilmark{2}}
\affil{California Institute of Technology, Pasadena, CA 91125, USA}
\email{sradford@submm.caltech.edu}



\begin{abstract}
  We report  single-dish multi-transition measurements of  the $^{12}$CO, HCN,
  and HCO$^+$  molecular line emission as well  as HNC J=1--0 and  HNCO in the
  two ultraluminous infra-red galaxies Arp\,220 and NGC\,6240.  Using this new
  molecular  line  inventory,  in   conjunction  with  existing  data  in  the
  literature, we compiled the most  extensive molecular line data sets to date
  for such  galaxies.  The many  rotational transitions, with  their different
  excitation requirements,  allow the study of  the molecular gas  over a wide
  range of  different densities and temperatures with  significant redundancy, 
  and thus allow good constraints on  the properties of the dense gas in these
  two systems.  The mass  ($\sim (1-2)\times 10^{10}\,M_{\odot}$) of dense
  gas  ($\ga 10^{5-6}\,  cm^{-3}$) found  accounts for  the bulk  of their
  molecular gas  mass, and  is consistent with  most of their  IR luminosities
  powered  by intense  star bursts  while self-regulated  by O,B  star cluster
  radiative pressure  onto the star-forming  dense molecular gas.   The highly
  excited HCN transitions  trace a gas phase $\sim  (10-100)\times$ denser than
that of the  sub-thermally excited HCO$^+$  lines (for both galaxies).  These
two phases are  consistent with an  underlying density-size power law  found for
Galactic GMCs (but  with a steeper exponent), with  HCN lines tracing denser and
more compact regions than  HCO$^+$. Whether this is true in IR-luminous, star
forming galaxies in general remains to be seen, and underlines the need for
observations of molecular transitions with high critical densities for a sample
of  bright (U)LIRGs  in the local  Universe --  a task for  which the HI-FI
instrument on board Herschel is ideally suited to do.  \end{abstract}




%
\section{Introduction}
A cardinal result in extra-galactic astronomy  over the last 20 years has been
the discovery by {\it IRAS} of  a population of local Ultra Luminous Infra-Red
Galaxies (ULIRGs -- Soifer et al.\ (1986); see also Sanders \& Mirabel (1996),
and Lonsdale, Farrah \& Smith (2006)). Although these galaxies are rare in the
local  Universe,   their  extreme  IR-luminosities  ($L_{\mbox{\tiny{IR}}}\sim
10^{12}\,\Lsolar$) and merger morphologies  strongly suggest that they signify
a highly active and important phase in galaxy evolution -- a scenario which is
corroborated  by the  rapid increase  in their  space density  with  redshift. 
Furthermore,  ULIRGs --  being  gas-rich mergers  harboring  the most  extreme
star-forming conditions  encountered in the  local Universe -- are  thought to
re-enact the galaxy formation processes we  are only barely able to glimpse in
the distant  Universe.  Although, detailed studies of  the interstellar medium
(ISM)  in proto-galaxies  at  high redshifts  will  be possible  with the  new
generation of  cm/mm/sub-mm telescopes and receivers coming  online within the
next decade, our expectations of how  such objects will look in mm/sub-mm line
emission  are  ill-informed  due  to  the  dearth of  similar  data  locally.  
Multi-line  studies of the  low-intensity starbursts  M\,82 and  NGC\,253 have
been made (Nguyen-Q-Rieu et al.\ 1992; Jackson et al.\ 1995; Paglione, Jackson
\& Ishizuki 1997),  but given that their star formation  rates are about three
orders of magnitude lower than that  of typical ULIRGs, they may very well not
be  good  guides for  the  state  of the  ISM  in  local  ULIRGs and  high-$z$
starbursts  (Aalto et  al.\ 1995;  Papadopoulos  \& Seaquist  1998; Downes  \&
Solomon 1998).

In contrast, local ULIRGs, owing  to their extreme IR/CO-brightness, are ideal
targets for  multi-line studies of heavy  rotor molecules such as  HCN, CS and
HCO$^+$.   These  lines are  usually  $5  - 30$  times  weaker  than their  CO
$J$-level  counterparts, making them  currently impossible  to detect  in more
distant galaxies  and/or galaxies with lower star  formation rates.  Secondly,
their  compact H$_2$  distributions  -- as  revealed  by interferometric  maps
(e.g.\ Scoville et al.\ 1997; Tacconi et al.\ 1999) -- allow us to collect the
total line  flux of  these often weak  lines with  a single pointing.  This is
important since multi-beam receivers, capable of observing many transitions of
a variety of  species within a reasonable amount of  telescope time, are still
lacking.  These  two factors, brightness and relatively  compact molecular gas
distributions, allow us  to study the properties of  the dense ($n(\mbox{H}_2)
\gs 10^4\,$cm$^{-3}$)  gas phase in local  ULIRGs in much  greater detail than
would otherwise be  possible.  This is of great  importance, because the dense
gas provides  the immediate  star formation fuel  in galaxies (Gao  \& Solomon
2004a,b;  Wu et al.\  2005).  The  possibility of  a universal  star formation
efficiency, as  proposed by these studies, can  now be checked in  a much more
powerful manner,  since we  can estimate the  dense H$_2$ mass  using multiple
lines rather than a single line such as HCN $J=1-0$.

\bigskip

In the local  Universe, Arp\,220 and NGC\,6240 are  ideally suited for studies
of  the dense gas  in extreme  starburst galaxies,  as they  are the  two most
studied  (U)LIRGs to  date.  They  have the  best sampled  FIR/sub-mm spectral
energy distributions  (SEDs) of their class  (e.g.\ Dopita et  al.\ 2005), and
the largest number  of mm/sub-mm CO, HCN, CS,  and HCO$^+$ transitions already
observed.  Thus not  only are their dust continuum  SEDs well constrained, but
it is also relatively straightforward to fill the gaps in their molecular line
inventory.  Being  the nearest  and best studied,  Arp\,220 is the  ULIRG most
frequently  used  as a  template  for  high-$z$  studies of  starbursts.   The
prominent Active Galactic Nucleus (AGN) present in NGC\,6240 (even double-AGN,
Komossa et  al.\ (2003),  but not  in Arp\,220, cf.\  Iwasawa et  al.\ (2005))
offers the possibility to study the  influence of an AGN on the molecular ISM,
for  example  how  X-ray   Dissociation  Regions  (XDRs)  affect  H$_2$  cloud
chemistry,  significantly altering  the HCO$^+$  and HCN  abundances  (Lepp \&
Dalgarno 1996; Usero et al.\ 2004).

In this  paper we  present single-dish observations  of a number  of molecular
emission lines from  Arp\,220 and NGC\,6240, which together  with results from
the  literature allows  us  to compile  a  very large  (and almost  identical)
molecular  transition catalog for  these two  extreme starbursts.   Armed with
this unique catalog, and the large range of excitation conditions it can probe
(with significant overlap/redudancy between HCN and CS lines) we use it to: a)
set constraints on  the physical properties of the  dense molecular gas phase,
b) find the  gas density probed per molecular species and  its mass range, and
c) explore  underlying star formation  efficiencies, density-size hierarchies,
and examine the species best suited for tracing the true star forming phase.

The paper is outlined as follows: observations and results are described in \S
\ref{section:observations}   and   \ref{section:results},  respectively.    \S
\ref{section:analysis}  details  the   analysis  of  our  observations,  which
includes deriving dense gas mass  and star formation efficiencies based on the
radiative    transfer    modeling   of    the    observed    lines.   In    \S
\ref{section:discussion} we  discuss the implications  of our findings  in the
context  of  upcoming  missions  such  as  Herschel.  In  the  flat  cosmology
($\Omega_m=0.27$,     $\Omega_{\Lambda}     =     0.73$,    and     $H_0     =
71$\,km\,s$^{-1}$\,Mpc$^{-3}$ -- Spergel et al.\ 2003) adopted throughout this
paper  the  luminosity  distances  to  Arp\,220 and  NGC\,6240  are  77.5  and
105.5\,Mpc, respectively.

\section{Observations}\label{section:observations}
The 15-m  James Clerk Maxwell  Telescope\footnote{The JCMT is operated  by the
  Joint Astronomy Centre on behalf  of the United Kingdom Particle Physics and
  Astronomy  Research  Council   (PPARC),  the  Netherlands  Organisation  for
  Scientific Research, and the National Research Council of Canada.} (JCMT) on
Mauna Kea, Hawaii  was used to observe a number  of $^{12}$CO, $^{13}$CO, HCN,
CS,    and    HCO$^+$    lines    in    Arp\,220    and    NGC\,6240    (Table
\ref{table:jcmt-observations}).    These  observations   were  done   in  good
($\tau_{225\,\mbox{\tiny{GHz}}}   \la    0.035$)   to   medium    ($0.08   \la
\tau_{225\,\mbox{\tiny{GHz}}}\la 0.12$) weather conditions during several runs
on the  JCMT between January 2003  and December 2004.  The  two JCMT receivers
$A3$(211-279\,GHz) and  $B3$(312-370\,GHz) were  employed in order  to observe
all the lines,  with the $B3$ receiver  tuned in SSB mode. In  order to ensure
baselines  as   flat  as  possible,   the  observations  were  done   in  fast
beam-switching  mode with a  chop throw  of $30\arcsecs$  (AZ) and  a chopping
frequency  of   1\,Hz.   As  backend  we  used   the  Digital  Autocorrelation
Spectrometer (DAS)  in its widest, $\sim  1.8$\,GHz ($\sim 1100$\,km\,s$^{-1}$
for the  B3 receiver) mode  to cover the  full velocity-width of  the expected
broad lines.  The pointing was checked every 1-2\,hr throughout each observing
run  and  was $\sim  3\parcsec5$  (rms) or  better.   Aperture  and main  beam
efficiencies were derived from observations  of Mars and Uranus throughout the
observing runs  and were compared to  mean JCMT values over  for the observing
periods.  There  is excellent agreement between  the two sets of  values so we
adopt  the latter under  the assumption  the mean  values carry  less inherent
uncertainty than single measurements. The details of the observations, such as
beam  sizes,  line  frequencies,   system  temperatures  and  total  on-source
integration times, are summarized in Table \ref{table:jcmt-observations}.

\bigskip

The  HCN 1--0  and  $^{13}$CO 2--1  lines  were observed  with  the IRAM  30-m
telescope on  Pico Veleta (Spain) in June  2006 as part of  a larger molecular
line survey  of LIRGs  (Papadopoulos et al.   2007a).  Each line  was observed
simultaneously in two  polarizations using a pair of  SIS receivers. The 3\,mm
A100/B100 receivers  were used for the  HCN 1--0 line in  conjunction with two
1\,MHz filter banks, each with 512  channels and a channel separation of $\sim
3.4\,$km\,s$^{-1}$  at the observed  frequency of  87\,GHz. For  the $^{13}$CO
2--1 line,  we used the 1\,mm  A230/B230 receivers together with  the two wide
4\,MHz filter  banks, each  with 1\,GHz bandpass  and a channel  separation of
$\sim 5.5$\,km\,s$^{-1}$  at the observed  frequency of 216.5\,GHz.   The beam
size  of  the 30-m  telescope  is 28\arcsecs  at  115\,GHz  and 12\arcsecs  at
217\,GHz. The data  were acquired with the New Control  System (NCS) in series
of  four  minute  scans,  each  comprised of  eight  30\,s  sub-scans.   Total
integration times were 1.2 hours for Arp\,220 and almost 2 hours for NGC\,6240
in  HCN 1--0  and  $^{13}$CO 2--1  (Table \ref{table:jcmt-observations}).   To
obtain  the  flattest  possible  baselines, the  wobbler  switching  (nutating
subreflector) observing mode was employed with a frequency of 0.5\,Hz and beam
throws of  $180-240\arcsecs$.  Both targets  were observed under  good weather
conditions in the late night hours  with $T_{sys} \sim 95-125$\,K at 3\,mm and
$225-500$\,K  at  1\,mm  on  the  antenna  temperature  ($T^{*}_{A}$)  scale.  
Pointing and  focus were checked  frequently throughout observations,  and the
pointing errors were found to be $\sim3\arcsecs$ (rms).

\bigskip

Finally, we  present previously unpublished observations  of Arp\,220 obtained
with the IRAM  30-m during the period 1988-1992.  These data include $^{13}$CO
1--0, CS  2--1, 3--2,  C$^{18}$O 1--0, and  HNC 1--0,  the isomer of  HCN. For
these  observations, SIS  receivers  were used  together  with the  $512\times
1\,\mbox{MHz}$  spectrometers.   The observations  were  done  with a  wobbler
switching of  180-240\arcsecs~in azimuth. The  data were calibrated  with cold
and ambient  loads and,  depending on frequency  and weather,  single sideband
temperatures were  200-600\,K. The pointing, checked with  planets and quasars
on a regular basis, was $\sim3\arcsecs$ (rms).

\bigskip

The spectra of  all the molecular lines observed in  Arp\,220 and NGC\,6240 as
part  of  this  study  are  shown  in  Fig.\  \ref{figure:spectra-arp220}  and
\ref{figure:spectra-ngc6240}, respectively.  The  spectra have been plotted on
the  $T_A^*$ temperature scale,  i.e., the  antenna temperature  corrected for
atmospheric  absorption and  rearward spillover  (Kutner \&  Ulich  1981). The
velocity-integrated line flux densities were estimated using
\begin{eqnarray}
I            &=& \int_{\Delta v} S_{\nu}\,dv = \frac{8k_B}{\eta_{ap} \pi D^2} K_c(x) \int_{\Delta v} T^*_{A}\,dv\\
             &=& \Gamma \eta_{ap}^{-1} K_c(x) \int_{\Delta v} T^*_{A}\,dv,  
\end{eqnarray}
where $\eta_{ap}$ is the aperture  efficiency, $D$ the telescope diameter, and
$k_B$ Boltzmann's constant  (Gordon, Baars \& Cocke 1992).   We find $\Gamma =
15.625\,\mbox{Jy\,K$^{-1}$}$     for     the     JCMT    and     $\Gamma     =
3.907\,\mbox{Jy\,K$^{-1}$}$  for  the  IRAM  30-m Telescope.  $T^*_A$  is  the
baseline subtracted spectrum,  and $K_c(x) = x^2/(1-e^{-x^2}), x=\theta_s/(1.2
\theta_{\mbox{\tiny{{\sc   hpbw}}}})$   (where   $\theta_s$=source   diameter)
accounts  for  the  coupling  of  the   gaussian  part  of  the  beam  with  a
finite-sized, disk-like source.   Interferometric observations have shown that
the  CO emitting regions  in Arp\,220  and NGC\,6240  are $\sim  4\arcsecs$ in
angular  diameter (Tacconi et  al.\ 1999;  Scoville et  al.\ 1997).   Our line
observations  collect, therefore,  all the  flux, with  minimal  dependence on
$K_c$,  which is  at most  $\sim  1.04$ for  the highest  frequency (and  thus
smallest beam) observations reported here.

All of the measured line fluxes along with their errors are tabulated in Table
\ref{table:lines}. The errors include  thermal and calibration errors, as well
as errors on  the assumed aperture efficiencies. Each of  these are assumed to
be uncorrelated and thus added in  quadrature. The latter two sources of error
were found to  contribute by no more than 10 per-cent  each, while the thermal
noise was calculated for each line using:
\begin{equation}
\sigma (\int_{\Delta v} T^*_A\,dv) = \sqrt{N_{\Delta v}} \left ( 1 + \frac{N_{\Delta v}}{N_{bas}}\right )^{1/2} 
\sigma (T^*_{A,ch}) \Delta v_{ch},
\end{equation}
where  $N_{\Delta v} (=  \Delta v/\Delta  v_{ch})$ is  the number  of channels
across the  line, $N_{bas}$  is the number  of channels (each  $\Delta v_{ch}$
wide) used to  estimate the baseline level, and $  \sigma (T^*_{A,ch})$ is the
channel-to-channel rms noise, assumed to be uniform across the spectrum.

\subsection{Calibration issues with the A3 receiver}
At $\nu  > 230\,$GHz the $A3$ receiver  on the JCMT may  present a calibration
problem that has  to do with the  fact that the sideband ratio  in this Double
Sideband (DSB)  receiver is not  unity but  can vary as  a function of  the LO
frequency.  This  may have  affected our HCN  3--2 measurements, which  we now
reduce separately  for each  sideband, and intend  to correct  properly before
co-adding.

In the standard calibration the  raw normalized spectrum is scaled by $T_{cal}
= (1+G_{is})\Delta T$, where $G_{is}=G_i/G_s$ ($=1$ for DSB mode) is the ratio
of the  image and signal  sideband gains, and  $\Delta T$ depends on  the mean
atmospheric  and  telescope  cabin  temperatures, the  telescope  transmission
efficiency, and  the line-of-sight  optical depth (see  Eq.  A12 of  Kutner \&
Ulich 1981). The  first step for reducing spectra for  which $G_{is}\neq 1$ is
to divide the spectra  by 2 in order to undo the  standard online calibration. 
Next, the  LSB and  USB tuned spectra  were averaged and  corrected separately
using    the   $G_{is}(\nu_    {LO})$   dependence    given   on    the   JCMT
website\footnote{{\it
    http://www.jach.hawaii.edu/JCMT/spectral\_line/Instrument\_homes/RxA3i/sidebands.html}},
which  for  for the  HCN  3--2  line yielded  $G_{is}  =  0.81$. Finally,  the
resulting corrected spectra were co-added.

\section{Results}\label{section:results}
In  this section  we describe  the new  transitions observed  in  Arp\,220 and
NGC\,6240 as  part of this  study, along with previously  detected transitions
that we  have re-observed.  In  order to compare  the line strengths  not only
between the  two galaxies  but also between  different transitions  within the
same system, we have  calculated velocity-integrated line luminosities for all
of the observed transitions. These were calculated using
\begin{equation}
L' = 3.25\times 10^7 \left (\frac{\nu_{obs}}{\mbox{\small{GHz}}}\right )^{-2}  \left ( \frac{D_L}{\mbox{\small{Mpc}}}\right )^2 (1+z)^{-3} \left ( \frac{\int_{\Delta v} S_{\nu} dv}{\mbox{\small{Jy\,km\,s$^{-1}$}}} \right ), 
\end{equation}
(Solomon, Downes \& Radford 1992b).
The resulting line luminosities calculated in this manner are tabulated in Table \ref{table:lines}.

\subsection{Arp\,220}
\begin{figure}
\plotone{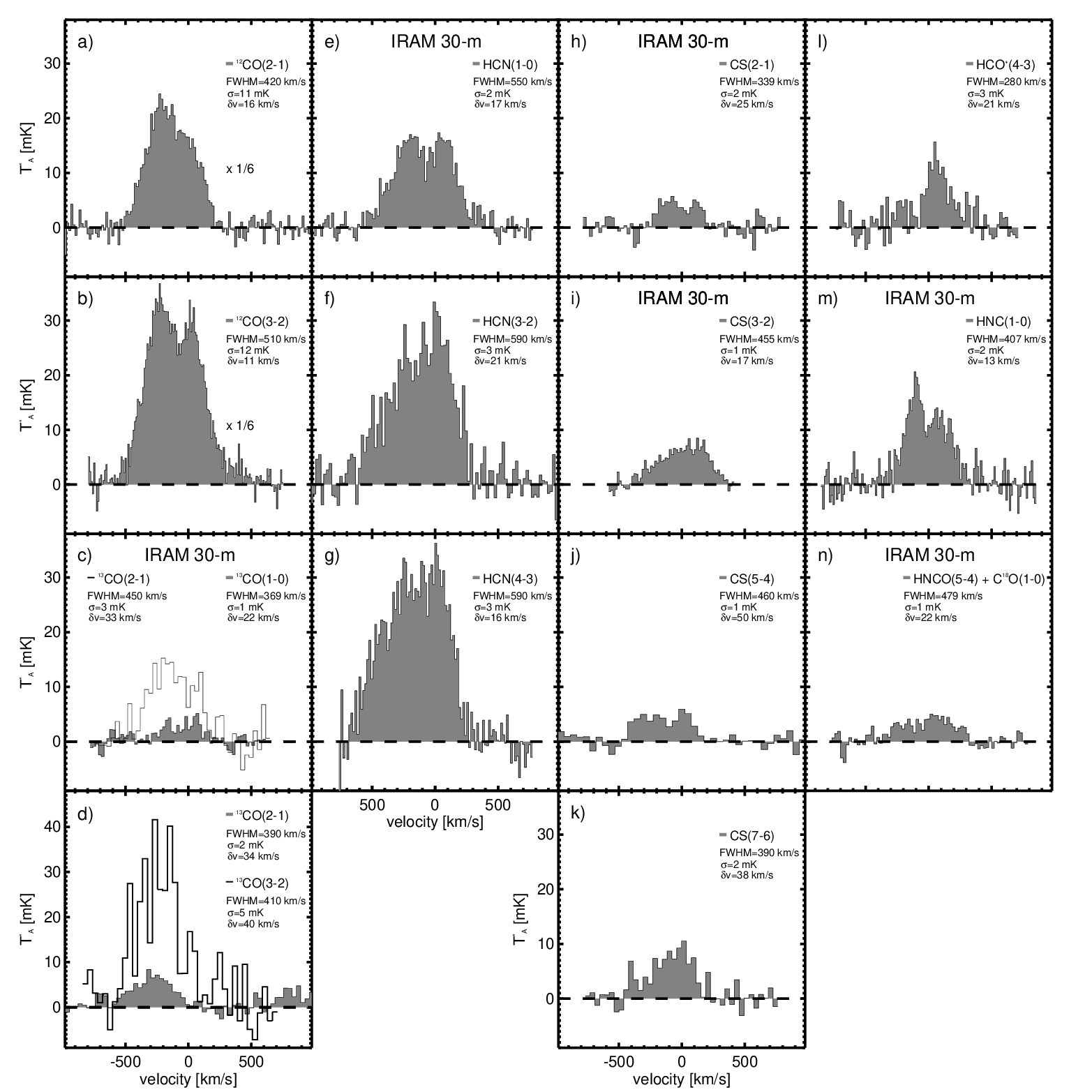}
\caption{Molecular line spectra observed in Arp\,220. The $^{12}$CO 2--1 and 3--2
spectra have been scaled down by a factor six in order to fit on the plot.
The velocity-axis is relative to $v_{\mbox{\tiny{LSR}}} = 5454\,$km\,s$^{-1}$.
The {\sc fwhm} line width and channel-to-channel rms noise are given for each line.
All spectra were obtained with the JCMT except for those in panel c), e), h)-i), and
m)-n) which are from the IRAM 30-m Telescope.  
}
\label{figure:spectra-arp220}
\end{figure}

In  Arp\,220, high-resolution interferometric  observations have  revealed the
double-horn profiles  of the $^{12}$CO 1--0  and 2--1 lines  (Scoville et al.\ 
1991) to originate  from two  nuclei embedded within  a circumnuclear  ring or
disk (Scoville  et al.\ 1997;  Downes \& Solomon  1998). We see the  same line
profile in  our $^{12}$CO 2--1  spectrum (Fig.\ \ref{figure:spectra-arp220}a),
with  a  strong emission  peak  at  $\simeq  -200$\,km\,s$^{-1}$ (relative  to
$V_{LSR}=5454\,\mbox{km}\,\mbox{s}^{-1}$) and  a less intense  peak at $\simeq
+50$\,km\,s$^{-1}$. These two peaks  correspond, respectively, to the western,
blueshifted nucleus  and to the  eastern, redshifted nucleus in  the molecular
disk (Downes \& Solomon  1998).  Our $^{12}$CO 3--2 line, single-dish spectrum
show a double-horn profile similar to the lower transitions (Lisenfeld et al.\
1996; Mauersberger et al.\ 1999).  

One subtle difference  exists however, and is confirmed by our own $^{12}$CO
3--2 spectrum (Fig.\ \ref{figure:spectra-arp220}b). The relative strength of the
peak at $\simeq +50$\,km\,s$^{-1}$ is greater in the 3--2 line than in  the 1--0
and 2--1  spectra. In the  3--2 spectrum the western component is only slightly
stronger than the eastern. Recent interferometric observations with the
Submillimeter Array (SMA) showed that the CO 3--2 and the 890-$\mu$m dust
emission is indeed strongest in the western nucleus (Sakamoto et al.\ 2008).

The $^{13}$CO lines  observed in Arp\,220 with the IRAM 30-m  and the JCMT are
shown in Fig.\ \ref{figure:spectra-arp220}c and d, respectively.  Although the
$^{13}$CO lines seem somewhat narrower than their $^{12}$CO counterparts, this
may be simply an effect of the low signal-to-noise ratio in the line wings and
the  uncertain baseline  level.  In  general all  the $^{12}$CO  and $^{13}$CO
lines  seem  to  have  very   similar  line  profiles,  within  the  available
signal-to-noise. Both  the $^{13}$CO 1--0  and 2--1 spectra observed  with the
IRAM  30-m  show tentative  evidence  of  double  peaks, consistent  with  the
$^{12}$CO lines. In the 1--0  spectrum, both peaks are equally strong, whereas
the  peak  at $\simeq  -200\,\rm{km}\,\rm{s}^{-1}$  is  stronger  in the  2--1
spectrum.   This also appears  to be  the case  in the  2--1 and  3--2 spectra
observed with the JCMT  (Fig.\ \ref{figure:spectra-arp220}d), although we note
the 3--2  spectrum in particular is  very noisy.  The fact  the $^{13}$CO 2--1
spectra obtained  with the two  telescopes are in  agreement in terms  of line
shape as well as flux density gives us confidence our observations are able to
robustly discern the line shapes of even faint lines like $^{13}$CO 2--1.

The HCN 1--0,  3--2, and 4--3 spectra observed in Arp\,220  are shown in Fig.\ 
\ref{figure:spectra-arp220}e-g,  respectively. The HCN  1--0 spectrum  shows a
clear  double-horn profile  similar  to $^{12}$CO  3--2,  with equally  strong
peaks. The HCN 3--2  and  4--3  line  profiles   are  both  very  broad  ({\sc
fwhm}~$\simeq 540-590$\,km\,s$^{-1}$)  and share  the same  overall shape.  They
both show evidence of  double peaks with the $\simeq  +50$\,km\,s$^{-1}$
component being more prominent. The  shape is unlike the HCN(1--0) spectrum --
as well as the $^{12}$CO  2--1 and  3--2  spectra --  where  the $\simeq
-200$\,km\,s$^{-1}$ component is brighter. 

The  CS lines  --  of which  only CS  3--2  has been  previously reported  for
Arp\,220 --  trace a much  overlaping excitation range  with that of  HCN (see
Table 6), and are shown in Fig.\ \ref{figure:spectra-arp220}h-k.  All CS lines
have double-horn  profiles, with the $\simeq  +50$\,km\,s$^{-1}$ peak stronger
in the  CS 3--2, 5--4, and  7--6 spectra, while  the two peaks are  almost the
same strength in the 2--1  spectrum.  This mimicks the observed evolution from
low- to  high-$J$ in the  HCN line profiles.   We note, however,  the apparent
asymmetry  of the CS  3--2 line  may be  an artifact  of the  limited receiver
bandwidth.  In determining  the baseline,  we forced  the right  edge  of this
spectrum  to zero,  so  our line  flux estimate  for  this line  is likely  an
underestimate.

Finally, in the last column  of Fig.\ \ref{figure:spectra-arp220}, we show the
HCO$^+$   4--3,  HNC   1--0,  and   the  blended   C$^{18}$O  1--0   and  HNCO
5$_{0,5}$--4$_{0,4}$ lines.   Similarly to the  HCN and CS lines,  the HCO$^+$
spectrum peaks at $\simeq +50$\,km\,s$^{-1}$, and shows some faint emission at
$\simeq  -200$\,km\,s$^{-1}$,  which  corresponds  to the  stronger  $^{12}$CO
feature. In this respect, the HCO$^+$ spectrum looks similar to CS 7--6, where
the  western component  is almost  completely quenched.  In stark  contrast to
this, the double-horn spectrum of  HNC 1--0 shows a remarkably strong emission
peak at $\simeq -200$\,km\,s$^{-1}$,  suggesting a substantial fraction of the
HNC is  concentrated in the western  nucleus.  The blended  C$^{18}$O 1--0 and
HNCO     5$_{0,5}$--4$_{0,4}$      spectrum     is     shown      in     Fig.\ 
\ref{figure:spectra-arp220}n.  In  order to determine  separate parameters for
C$^{18}$O 1--0 and HNCO 5$_{0,5}$--4$_{0,4}$, two gaussian components with the
same width and fixed separation were fitted. From the resulting intensities we
calculated integrated line fluxes and luminosities (Table \ref{table:lines}).

It is  telling that  the double-horn structure  seen in the  low-$J$ $^{12}$CO
spectra is also  discernible in the high-$J$ spectra of  HCN, CS, and HCO$^+$,
albeit with  the relative  strengths of  the two peaks  reversed.  In  fact, a
rather  clear  trend is  apparent:  the  $\simeq  -200$\,km\,s$^{-1}$ peak  is
stronger in the $^{12}$CO 2--1 spectrum, the peaks are about equally strong in
the $^{12}$CO 3--2  and HCN 1--0 profiles, and  the $\simeq +50$\,km\,s$^{-1}$
peak is  stronger in the  HCN 3--2  and 4--3 spectra.   Thus, as we  climb the
$J$-ladder and probe higher and higher critical densities, the strength of the
emission  shifts from the  western to  the eastern  nucleus.  This  picture is
confirmed by  the CS  7--6, and  HCO$^+$ 4--3 spectra  which have  even higher
critical densities than the HCN 1--0  line and are completely dominated by the
$\simeq +50$\,km\,s$^{-1}$ component with very little or no emission at all at
$\simeq -200$\,km\,s$^{-1}$. 

An important point to make here is that it is only due to the high-density gas
tracers such as HCO$^+$, HCN and CS that we are able to conclude that the
eastern nucleus harbours most of the dense gas. Without those, we would have
been inclined to conclude, based on the CO 3--2 spectrum (Fig.\
\ref{figure:spectra-arp220}b) and its distribution (Sakamoto et al.\ 2008), that
most of the dense gas was contained in the western component.

Given the modest (yet real)  differences of the relative line strengths across
the  velocity   range  of  Arp\,220,   and  for  reasons  of   simplicity  and
straightforward  comparison   to  NGC\,6240,  we  decided  to   use  the  line
intensities average over its entire profile and the associated line ratios in
our susbesquent analysis (\S \ref{section:analysis}).

\subsection{NGC\,6240}
\begin{figure}
\plotone{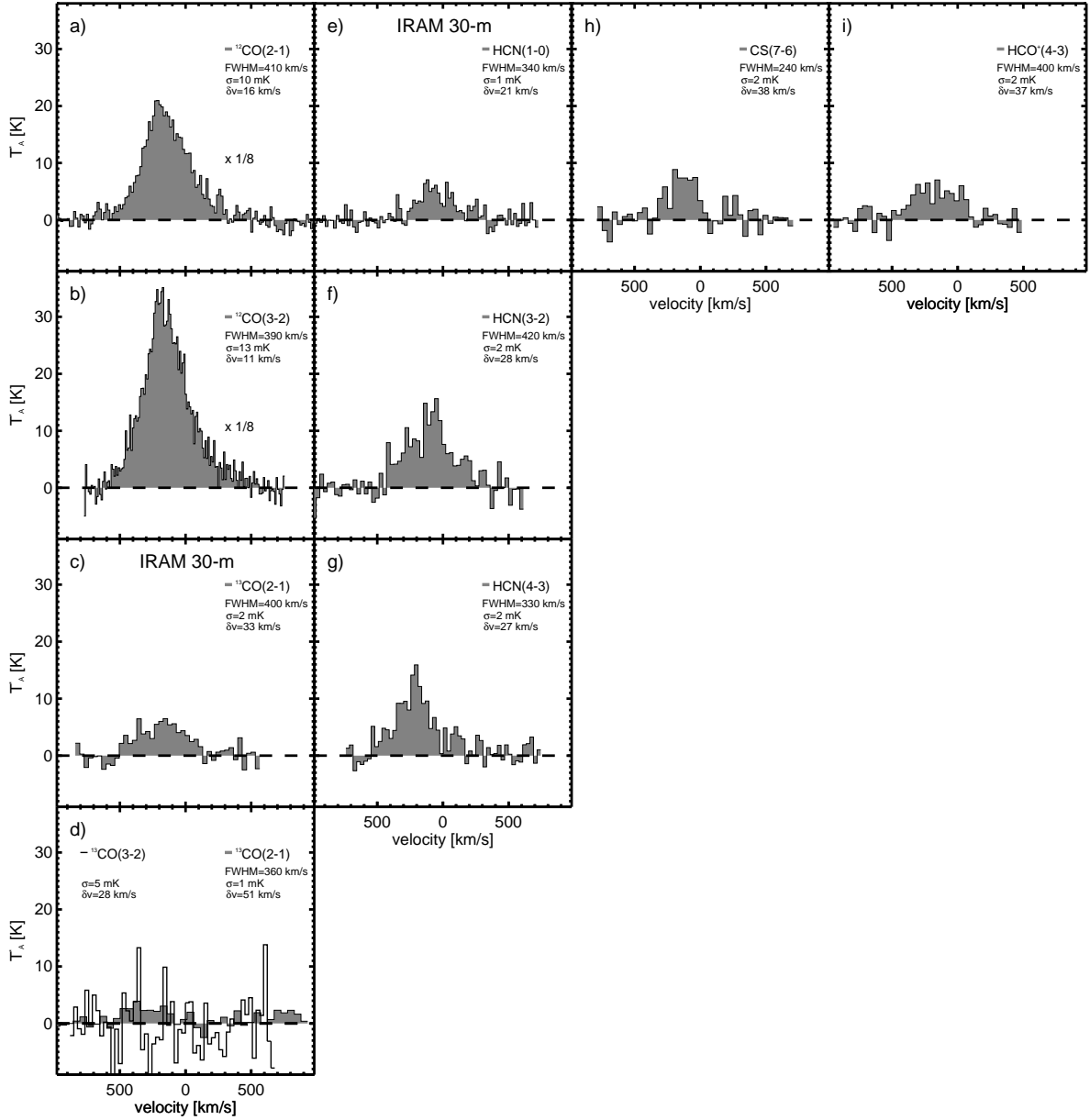}
\caption{Molecular line spectra observed in NGC\,6240. 
The velocity-axis is relative to $v_{\mbox{\tiny{LSR}}} = 7359\,$km\,s$^{-1}$,
and the $^{12}$CO 2--1 and 3--2 spectra have been scaled down by
a factor eight. All spectra were obtained with the JCMT except for those in panel c) and e) 
which are from the IRAM 30-m Telescope.  
Other details are the same as in Fig.\ \ref{figure:spectra-arp220}.
}
\label{figure:spectra-ngc6240}
\end{figure}

The  $^{12}$CO  2--1  and  3--2  spectra  of  NGC\,6240  are  shown  in  Fig.\ 
\ref{figure:spectra-ngc6240}a  and  b,  respectively.   Unlike  Arp\,220,  the
$^{12}$CO 3--2  and 2--1  spectra in NGC\,6240  have almost  identical shapes,
with spectra having a single-peaked  profile, skewed towards the blue ($\simeq
-200$\,km\,s$^{-1}$  relative  to  $v_{LSR}=7359\,\mbox{km}\,\mbox{s}^{-1}$).  
This  is  consistent  with  high-resolution  interferometric  observations  of
low-$J$\, $^{12}$CO  lines in  the inner regions  of NGC\,6240,  which suggest
more than half  of the molecular gas resides in  a central structure dominated
by turbulent  flows, rather  than systematic orbital  motion (Tacconi  et al.\ 
1999).  The  $^{13}$CO 2--1 spectra obtained  with the JCMT and  IRAM 30-m are
shown  in Fig.\  \ref{figure:spectra-ngc6240}c and  d, respectively,  and both
appear similar  in shape to  their $^{12}$CO counterparts.   Unfortunately, we
cannot      compare      the      $^{13}$CO     3--2      spectrum      (Fig.\ 
\ref{figure:spectra-ngc6240}d) as no obvious  detection of this transition was
made.  We are able, however, to  set a sensitive upper limit on its integrated
flux (Table \ref{table:lines}).

The HCN  1--0, 3--2, and 4--3  transitions observed in NGC\,6240  are shown in
Fig.\  \ref{figure:spectra-ngc6240}e-g.   Although   the  HCN  1--0  line  has
previously been  detected (Solomon, Downes  \& Radford 1992a; Tacconi  et al.\ 
1999;  Nakanishi et  al.\ 2005),  this is  the first  time the  3--2  and 4--3
transitions have  been detected in this  galaxy.  All three HCN  lines peak at
$\simeq -200$\,km\,s$^{-1}$ and are consistent with the CO line profiles.

Unlike Arp\,220, where several CS lines  were observed, only the 7--6 line was
observed in NGC\,6240 due to limited  observing time. The spectrum is shown in
Fig.\  \ref{figure:spectra-ngc6240}h. This  is  the only  detection  of CS  in
NGC\,6240 to date and the line profile is consistent with those of CO and HCN.

Finally, the  HCO$^+$ 4--3  spectrum observed in  NGC\,6240 is shown  in Fig.\ 
\ref{figure:spectra-ngc6240}i.  The  HCO$^+$  4--3  line profile  is  in  good
agreement with the 1--0 line (Nakanishi et al.\ 2005) and is consistent with a
single peak at $\simeq -200\,$km\,s$^{-1}$.

In summary,  all of the transitions  detected in NGC\,6240  exhibit a similar,
Gaussian-like velocity  profile, peaking at  about the same  velocity ($\simeq
-200\,$km\,s$^{-1}$), in  contrast to  Arp\,220 where all  the lines  not only
show a  double-peaked profile but a  strong evolution in the  dominant peak is
also discernible.   The similar  velocity profile over  such a large  range of
densities probed in  NGC\,6240 signifies a kinematical state  of the molecular
gas where  the various  phases remain well-mixed  down to much  smaller scales
than in Arp\,220 where the  denser phases seem settled to two counter-rotating
dense  gas disks,  enveloped  by a  much  more diffuse  CO-bright  gas phase.  
Sub-arcsec interferometric imaging of the HCN, CS and HCO$^+$ imaging of these
systems is  valuable in confirming  such a picture,  a possible result  of the
different merger state of these two strongly interacting systems.

\subsection{Literature data}
A major goal of this study was to use our observations as well as results from
the  literature to  compile  a  large, identical  molecular  line catalog  for
Arp\,220  and NGC\,6240.   Thus the  literature was  scanned for  all relevant
molecular line  measurements and for  each transition the weighted  average of
every  independent  flux  measurement   was  calculated.   In  cases  where  a
measurement deviated significantly from  the mean for obvious reasons (limited
bandwidth, poor  conditions, etc.), it  was discarded from the  average.  Note
that often in the literature line measurements are given only in (main beam or
antenna) temperature units and only rarely in Janskys. In those cases, we used
the aperture  and/or main beam efficiencies  quoted in the paper  or, if those
were not  given, the  average values for  the particular  telescope (typically
tabulated in the observing manual). The final tally of lines is given in Table
\ref{table:lines}  along  with  their  corresponding fluxes  and  global  line
luminosities.   Global  line  ratios  were  calculated  as  the  velocity/area
averaged brightness temperature ratios (equivalent  to taking the ratio of the
line  luminosities) and are  listed in  Tables \ref{table:line-ratios-diffuse}
and \ref{table:line-ratios-dense}.

Our  observations of  $^{12}$CO  2--1  in both  Arp\,220  and NGC\,6240  yield
somewhat  higher  line fluxes  than  previous  single-dish and  interferometry
observations listed in Table \ref{table:lines}, although, we note that Wiedner
et al.\  (2002) using the JCMT measured  an even larger line  flux in Arp\,220
than ours.  A careful scrutiny of  the individual sub-scans in our data and of
the calibration on  the dates the observations were  made revealed nothing out
of the  ordinary. In fact,  given that each  source was observed  on different
nights, yet  both have $\sim 40$  per cent higher line  fluxes than previously
published values, suggest that the offset may well be real. A close inspection
of previously  published $^{12}$CO 2--1  spectra reveals they all  suffer from
either limited  bandwidth (the case for single-dish  measurements), and/or the
sources  have been  over-resolved spatially  (in the  case  of interferometric
observations).  Because only few  observations favor the higher $^{12}$CO 2--1
fluxes in  Arp\,220 and NGC\,6240,  however, we conservatively adopt  the mean
flux value of all available measurements  for our analysis.  The HCN 4--3 line
in Arp\,220  was first observed  by Wiedner et  al.\ (2002), who found  a line
flux only half as large as ours.  Here, we adopt our own measurement since our
spectrum  has a  higher signal-to-noise  ratio than  that of  Wiedner  et al.\ 
(2002), and the latter  used aperture efficiencies which deviate significantly
from the standard values given for JCMT.

In NGC\,6240  we find a  HCN 1--0 flux  that is half the  previously published
single-dish measurement (Solomon, Downes \& Radford 1992a) but is in excellent
agreement with  the infererometric  measurement by Nakanishi  et al.\  (2005). 
Because  our data  were obtained  over  the course  of two  days in  excellent
weather conditions and  because a careful reduction of the  data from each day
separately yielded consistent  results, we find the lower  values more likely. 
In the  case of Arp\,220, our estimate of the HCN 1--0 line flux is about
35 per cent larger than that of Solomon, Downes \& Radford (1992a), which is
just within the errors.
Recent, independent observations of the HCN 1--0 line in Arp\,220 and NGC\,6240 with the IRAM 30-m
Telescope   yield   line   fluxes   in   agreement   with   our   measurements
(Graci\'{a}-Carpio  private  communications),  suggesting  the HCN  1--0  line
fluxes given by  Solomon, Downes \& Radford (1992a) may  have been affected by
systematic calibration uncertainties. Nonetheless, for the line ratio analysis
in this paper,  we conservatively adopt the weighted average  of all three HCN
1--0 measurements in these two systems.

\bigskip

We end  this section  by noting that  Table \ref{table:lines}  constitutes the
most extensive molecular  line catalog available for any  (U)LIRG. In terms of
extragalactic objects in general, it is matched only by data published for the
much less extreme starburst galaxies M\,82 and NGC\,253 (e.g., Wei\ss~ et al.\ 
2005;  Seaquist et al.\  2006; Bayet  et al.\  2004; Bradford  et al.\  2003). 
However, unlike  M\,82 and  NGC\,253, where only  the $^{12}$CO 1--0  and 2--1
transitions have been observed over  their entire extent (e.g.\ Wei\ss, Walter
\& Scoville  2005), our study picks  up the global emission  from Arp\,220 and
NGC\,6240  for  all  transitions,  allowing  for  an  unbiased  study  of  the
properties  of   {\it  their  entire   molecular  gas  reservoir}   using  the
corresponding  mm/sub-mm  line  ratios.   This  mirrors  observations  of  the
mm/sub-mm lines  observed in  high-$z$ starbursts where  much fewer  lines are
available, and thus  cannot yield meaningful constraints on  the conditions or
the mass per gas phase for their molecular gas reservoirs.

\section{Analysis}\label{section:analysis}
The first indication  that the molecular ISM in  extreme starburst galaxies is
fundamentally different from that of normal galaxies came from the observation
that the former have much larger $^{12}$CO$/^{13}$CO line ratios (e.g.\ Casoli,
Dupraz \& Combes 1992).   These are  almost exclusively  observed in  ULIRGs  and were
interpreted as either  due to missing $^{13}$CO caused  by a top-heavy initial
mass function in these extreme systems  (Casoli, Dupraz \& Combes 1992) or due to purely
optical depth  effects (i.e., low $^{12}$CO  optical depths) caused  by a warm
and  highly turbulent H$_2$  medium present  in these  objects (Aalto  et al.\ 
1995).   Then  the  bulk of  the  $^{12}$CO  emission  arises from  a  diffuse
enveloping phase  where $^{12}$CO  has low to  moderate optical  depths ($\tau
\sim 1$), and the large $^{12}$CO/$^{13}$CO ratios are due to $^{13}$CO-bright
gas with  a much a smaller  filling factor than the  diffuse $ ^{12}$CO-bright
phase.

It is  instructive to examine how  far a single gas  phase can go  in terms of
explaining the observed  line ratios in Arp\,220 and  NGC\,6240 before the fit
becomes inadequate and a second gas  phase must be introduced.  To this end we
employ a Large  Velocity Gradient (LVG) model based on  the work by Richardson
(1985), which  uses the observed CO  line ratios to constrain  the gas density
($n(\mbox{H}_2)$),        kinetic        temperature       ($T_k$),        and
$\Lambda_{\mbox{\tiny{CO}}}  =  [\mbox{CO}/\mbox{H}_2](dv/dr)^{-1}$, which  is
the CO  abundance relative to  H$_2$ divided by  the cloud velocity  gradient. 
The models  span $n(\mbox{H}_2) = 10^2  - 10^9$\,cm$^{-3}$ (in  steps of $\log
n(\mbox{H}_2)  =  0.5$),  $T_k  =  10  -  100$\,K  (in  steps  of  5\,K),  and
$\Lambda_{\mbox{\tiny{CO}}}=10^{-6}-10^{-3}\,(\mbox{km}\,\mbox{s}^{-1}\,\mbox{pc}^{-1})^{-1}$
(in  steps of  $\log \Lambda_{\mbox{\tiny{CO}}}=0.5$).   The model  adopts the
CO-H$_2$ collisional rate constants from Flower \& Launay (1985), and included
all rotational levels up to $J=11$ for kinetic temperatures of 10, 20, 40, 60,
and  100\,K   were  included.    Collision  rates  for   intermediate  kinetic
temperatures were  estimated by  interpolation.  Finally, because  the assumed
$^{12}$CO$/^{13}$CO  abundance  ratio  can  be  important for  whether  a  fit
converges or  not, we considered models  with abundance ratios of  40, 60, and
100, values that span those found in the Milky Way (Langer \& Penzias 1993).

The need  for a two-phase molecular  ISM is perhaps most  obvious in Arp\,220,
where no  single LVG solution can  account for even the  $^{12}$CO 1--0, 2--1,
and 3--2  lines. In this  system we have  the interesting situation  where the
ratio $r_{21} =^{~12}$CO (2--1)/(1--0)$=0.7\pm 0.1$ points towards sub-thermal
excitation of even  $^{12}$CO 2--1 ($n_{crit}=2.7\times 10^3\,\mbox{cm}^{-3}$)
where typically  $n(\mbox{H}_2) \simeq 3\times  10^2\,\mbox{cm}^{-3}$ and $T_k
\simeq 40-60\,$K  fits such  a ratio well  (for LTE  such a low  ratio implies
$T_{k}\ls 7$\,K  the minimum gas temperature permitted  by cosmic-ray heating,
Goldsmith  \& Langer  1978).  Such  conditions cannot  reproduce  the observed
$r_{32} =^{~12}$CO(3--2)/(1--0)$=1.0\pm 0.1$ which indicates much more excited
gas with  density $\gs  3\times 10^3\,\mbox{cm}^{-3}$ and  kinetic temperature
$\gs 30$\,K.   Going ahead  to searcg  for a LVG  solution constrained  by the
$^{12}$CO 1--0,  2--1, 3--2 and just  one $^{13}$CO line  strength, the J=1--0
lines, yields predictably a much worse solution given by $n(\mbox{H}_2) \simeq
1-3\times 10^2\,\mbox{cm}^{-3}$ and $T_k  \simeq 35-45\,$K.  Aside from poorly
reproducing  the  aforementioned  line  ratios,  barely  satisfying  the  
$T_{k}\geq T_{dust}$ criterion  for these galaxies (see 4.2),  such a gas phase
corresponds to  brightness temperature $^{13}$CO  (2--1)/(1--0), (3--2)/(1--0)
ratios  of $\rm  r_{21}(^{13}CO)\sim  1.4-1.5$ and  $\rm r_{32}(  ^{13}CO)\sim
0.6-0.8$, in sharp  contrast to the observed ones of $\sim  1.64$ and $\sim 5$
respectively.

In the  case of  NGC\,6240, both $r_{21}=1.2\pm  0.2$ and  $r_{32}=1.1\pm 0.2$
ratios  indicate  higly  excited  gas  with  a  typical  gas  phase  model  of
$n(\mbox{H}_2)\simeq1\times  10^4\,\mbox{cm}^{_3}$ and $T_k  \simeq 40-50\,$K,
that can also fit the observed  large J=1--0 $ ^{12}$CO/$ ^{13}$CO ratio ($\rm
\mathcal{R}_{10}(model)=37$).   Nevertheless this solution  fails dramatically
for the J=2--1, J=3--2 $ ^{12}$CO/$  ^{13}$CO line ratios where it yields $\rm
\mathcal{R}_{21}\sim  \mathcal{R}_{32}\sim 17$, much  lower than  the observed
values (Table \ref{table:line-ratios-diffuse}), even for the largest plausible
abundance  of  [$^{12}$CO$/^{13}$CO]$=100$. This  could  not  have been  found
without our sensitive  $^{13}$CO 2--1 and 3--2 measurements,  which in the case
of NGC\,6240, have pushed the $\mathcal{R}_{21}$ and $\mathcal{R}_{32}$ ratios
to   the   highest   such    values   reported   in   the   literature. 

Finally,  using different $\rm  [ ^{12}CO/  ^{13}CO]$ abundance  ratios namely
$\rm [ ^{12}CO/ ^{13}CO]= 40, 60,  100$, encompasing the range deduced for the
Milky Way  (Langer \& Penzias 1993;  Casassus, Stahl \& Wilson  2005) we still
find no  single set  of average molecular  gas physical conditions  that comes
even close in reproducing the $  ^{12}$CO {\it and} the $ ^{13}$CO line ratios
in either galaxy.

\subsection{C$^{18}$O emission and the opacity of the $^{13}$CO lines}

The  emergence of $^{13}$CO  transitions as  the first  ones marking  a strong
deviation from any single gas  phase reproducing the relative strengths of the
$ ^{12}$CO transitions  is further highlighted in the case  of Arp\,220 by the
observed strong C$ ^{18}$O J=1--0 line emission. There the C$^{18}$O$/^{13}$CO
1--0 line  ratio of $1.0\pm  0.3$ is the  highest such ratio obtained  over an
entire galaxy --  -- by comparison, Papadopoulos, Seaquist  \& Scoville (1996)
measure  such  high  values towards  only  a  few  of  the brightest  GMCs  in
NGC\,1068.          Assuming        the        low         abundance        of
$[\mbox{C}^{18}\mbox{O}/^{13}\mbox{CO}]=0.15$ found for  the Milky Way (Langer
\&  Penzias 1993) such  high C$^{18}$O$/^{13}$CO  brightness ratios  argue for
much of the  $ ^{13}$CO J=1--0 emission emerging from a  phase where $\rm \tau
_{10}(  ^{13}CO)> 1$,  sharply  deviating from  the  optical thin  transitions
deduced from the LVG of only  the $ ^{12}$CO, $ ^{13}$CO transitions described
in  the previous  section. A  much denser  gas phase  can naturally  yield the
significantly   higher   optical   depths   needed   to   explain   the   high
C$^{18}$O$/^{13}$CO J=1--0 ratio.

\subsection{The properties of the dense molecular gas}\label{section:dense-phase}
In this  section we use the  many transitions of  HCN, CS, and HCO$^+$  in our
catalog  --  probing  a  wide  range  of  critical  densities  and  excitation
temperatures  in a redundant  manner (see  Table \ref{table:line-data})  -- to
constrain the properties of the dense gas phase.

In order  to model  the radiative  transfer of these  lines, we  constructed a
modified  version   of  the  original   LVG  code.   The   modifications  were
straightforward since  HCN, CS, and  HCO$^+$ are simple linear  molecules like
CO.  For HCN, CS, and HCO$^+$,  the collision rate constants provided by Green
\& Thaddeus (1974), Turner et al.\ (1992), and Flower (1999) were adopted.  We
explored a similar density  range as we did for the CO  lines, but limited the
temperature range  to $T_d \le  T_k \le 120\,$K.   The lower limit set  by the
dust  temperature  ($T_d$),  reflects the  fact  that  the  dust and  gas  are
thermally  de-coupled, with  the FUV-induced  photoelectric and  turbulent gas
heating (and  its cooling  via spectral line  rather than  continuum emission)
setting $T_k  \gs T_d$. Only in  the densest, most  FUV-shielded and quiescent
regions of the  GMCs can we expect  $T_k \simeq T_d$.  The upper  limit is the
typical equilibrium temperature of purely  atomic, C$^+$-cooled HI in the cold
neutral medium (e.g.\ Wolfire et al.\  2003).  Based on detailed fits to their
FIR/sub-mm SEDs (Dopita et al.\ 2005),  we adopted a dust temperature of $\sim
40$\,K for  both Arp\,220 and  NGC\,6240.  Since the molecules  have different
abundances, the  range in  $\Lambda$ is also  different for a  common velocity
gradient   ($dv/dr$).    In   our   own   Galaxy,   typical   abundances   are
$[\mbox{HCN}/\mbox{H}_2]  \sim 2\times  10^{-8}$ (Bergin,  Snell  \& Goldsmith
1996;  Lahuis \&  van  Dishoeck 2000),  $[\mbox{CS}/\mbox{H}_2] \sim  10^{-9}$
(Paglione et al.\ 1995;  Shirley et al.\ 2003) and $[\mbox{HCO}^+/\mbox{H}_2]=
8  \times 10^{-9}$  (Jansen 1995).   In order  to accommodate  the substantial
uncertainties associated with the abundances,  the range of $\Lambda$ used was
$10^{-10}-10^{-7}$,   $10^{-11}-10^{-8}$,  and  $10^{-11}-10^{-8}\,   \left  (
  \mbox{km\,s}^{-1}\mbox{pc}^{-1}  \right )^{-1}$  for HCN,  CS,  and HCO$^+$,
respectively. These correspond to velocity-gradients  in the range $\sim 0.2 -
800$\,km\,s$^{-1}$\,pc$^{-1}$ for the abundances quoted above.

In  the case  of Arp\,220,  the observed  HCN ratios  are consistent  with LVG
solutions     in      the     range     $n(\mbox{H}_2)\simeq     (0.3-1)\times
10^6\,\mbox{cm}^{-3}$,  and  $\Lambda_{\mbox{\tiny{HCN}}}\simeq  (0.3-1)\times
10^{-9}\,  \left ( \mbox{km\,s}^{-1}\mbox{pc}^{-1}  \right )^{-1}$,  for which
$\tau_{\mbox{\tiny{HCN(1-0)}}} \simeq 3-7$.   From the fit of the  CS lines we
find    $n(\mbox{H}_2)\simeq    (0.3-3)\times    10^6\,\mbox{cm}^{-3}$,    and
$\Lambda_{\mbox{\tiny{CS}}}\simeq    (0.01-1)\times     10^{-8}\,    \left    (
  \mbox{km\,s}^{-1}\mbox{pc}^{-1}  \right  )^{-1}$.   Finally, the  subthermal
HCO$^+$  ratios   allow  for  solutions  in   the  range  $n(\mbox{H}_2)\simeq
(0.3-1)\times                    10^4\,\mbox{cm}^{-3}$,                    and
$\Lambda_{\mbox{\tiny{HCO$^+$}}}\simeq   (0.3-1)\times   10^{-8}\,   \left   (
  \mbox{km\,s}^{-1}\mbox{pc}^{-1}  \right  )^{-1}$.   All  the  aforementioned
solutions are for  $T_{k}\geq 40\,$K and their details  are shown in Table
7.  There it  can be seen that  while HCN, CS,  and HCO$^+$ all trace  what in
broad  terms can  be  characterized  as dense  gas  (i.e., $n(\mbox{H}_2)  \gs
10^4\,\mbox{cm}^{-3}$), with  the density  range of the  CS and  HCN solutions
being very similar.  Moreover {\it we find the CS and HCN emission tracing gas
  with  densities $\sim 100  \times$ higher  than that  traced by  the HCO$^+$
  lines.}  This  simply reflects  the high  HCN, CS and  the low  HCO$^+$ line
ratios measured  in Arp\,220, further reinforced  by the $\sim  5$ times lower
critical densities  of the HCO$^+$  transitions than e.g  those of HCN  at the
same rotational level  (see Table 6).  These results  are certainly consistent
with  Galactic surveys of  CS lines,  which have  shown they  are unmistakable
markers  of very  dense, high-mass  star forming  cores (Plume  et  al.\ 1997;
Shirley et al.\ 2003).

For NGC\,6240, the  HCN LVG  fits converge  mostly  over $n(\mbox{H}_2)\simeq
(1-3)\times     10^5\,\mbox{cm}^{-3}$,      $T_k     \ge     40\,$K,     and
$\Lambda_{\mbox{\tiny{HCN}}}\simeq 3\times 10^{-10} - 1\times 10^{-8}\, \left (
  \mbox{km\,s}^{-1}\mbox{pc}^{-1} \right )^{-1}$,  with an optically thick HCN
J=1--0  line ($\tau_{\mbox{\tiny{HCN(1-0)}}}\simeq  3-29$).  From  the HCO$^+$
lines  we  find  solutions  in the  range  $n(\mbox{H}_2)\simeq  (0.3-1)\times
10^4\,\mbox{cm}^{-3}$,          $T_k          \ge         40\,$K,          and
$\Lambda_{\mbox{\tiny{HCO$^+$}}}\simeq   (0.3-1)\times   10^{-8}\,   \left   (
  \mbox{km\,s}^{-1}\mbox{pc}^{-1}  \right )^{-1}$.  Thus  in NGC\,6240,  as in
Arp\,220, HCN is tracing gas which is significantly denser than that traced by
HCO$^+$, over a similar range of temperatures.

In an attempt to ``break'' the significant degeneracy that still exists in the
LVG solutions for the  dense gas we make use of the fact  that at least in the
Galaxy, most of the dense  star-forming gas is found in self-gravitating cores
(Shirley et al.\  2003).  In the cases of Arp\,220 and  NGC\,6240, we use this
to try  and limit  the allowed ($n(\mbox{H}_2)$,  $T_k$, $\Lambda_{\mbox{\tiny
    X}}$)-parameter  space by  examining the  ratio of  the  velocity gradient
infered from the LVG fits to that expected for virialized cores, namely
\begin{equation}
K_{vir} = \frac{(dv/dr)_{\mbox{\tiny{$LVG$}}}}{(dv/dr)_{vir}} \sim 1.54 
\frac{[\mbox{X}/\mbox{H}_2]}{\sqrt{\alpha} \Lambda_{\mbox{\tiny{X}}}} 
\left (\frac{n(\mbox{H}_2)}{10^3\,\mbox{cm}^{-3}} \right ) ^{-1/2},
\label{equation:Kvir}
\end{equation}
where  [X$/$H$_2$]  is  the abundance  ratio  of  the  given molecule  X  (see
Papadopoulos  \& Seaquist  1999,  or  Goldsmith 2001  for  a derivation),  and
$\alpha\sim 1-2.5$  is a constant whose  value depends on  the assumed density
profile of a typical cloud (Bryant \& Scoville 1996).

For virialized gas we expect $K_{vir}\sim 1$  to within a factor of a few (due
to uncertainties in cloud geometry,  density profile, and abundance assumed to
deduce the observed $dv/dr$ from $\Lambda _{\mbox{\tiny{X}}}$).  LVG solutions that correspond
to $K_{vir}>>1$  indicate non-virial,  unbound motions and  are ruled  out for
dense star-forming gas along with  those that have $K_{vir}\ll 1$ (gas motions
cannot be slower  than those dictated by its  self-gravity).  For Arp\,220 and
Table  \ref{table:solutions}  we  see  that  for both  HCO$^+$  and  HCN,  LVG
solutions with $T_k\ge 40\,$K have $K_{vir}\simeq 1$, and thus this conditions
does not  single out any  particular range of  solutions.  For CS  however, we
find  $K_{vir}\ll 1$  for  most  solutions except  for  $T_k=35-40\,$K,  $\rm
n(\mbox{H}_2)=3\times 10^6\,cm^{-3}$       and       $\rm       \Lambda       _{\mbox{\tiny{CS}}}=
10^{-10}\,(km\,s^{-1}\,pc^{-1})^{-1}$.   {\it This  suggests  that the  true CS
abundance may  be significantly higher than  assumed (by at least  an order of
magnitude).}  Finally for  NGC\,6240 all solutions that fit  the HCO$^+$ ratios
are  consistent with  $K_{vir}\sim 1$,  while  for HCN  this is  the case  for
$T_k=40-55\,$K     and      $     n(\mbox{H}_2)=10^5\,$cm$^{-3}$     and     $\Lambda
_{\mbox{\tiny{HCN}}}=10^{-8}\,($km$\,$s$^{-1}\,$pc$^{-1})^{-1}$ (see Table 7).

\subsubsection{HNC vs.\ HCN}
Observations of HCN and HNC in Galactic GMCs (Goldsmith et al.\ 1981), as well
as steady state chemical models (Schilke et al.\ 1992), predict an increase in
the  HCN/HNC abundance  ratio  with increasing  temperature  and density,  and
attribute this  to the  destruction of HNC  (but not HCN)  via neutral-neutral
reactions  with   atomic  oxygen  or  hydrogen.   Then,   at  densities  $\sim
10^{4-5}\,\mbox{cm}^{-1}$,   temperature-independent   ion-neutral  reactions,
which form HCN and HNC in  equal amounts, may become dominant (Schilke et al.\ 
1992;  Aalto  et  al.\ 2002),  and  as  a  result  the  HNC abundance  can  be
substantial. However, given the high optical depths implied for the HCN J=1--0
transition  from our excitation  analysis of  the HCN  line ratios,  no useful
limits can be placed on the  $\rm [HNC/HCN]$ abundance ratio from the observed
brightness  temperarure  ratio  of  $1.2\pm  0.2$ measured  for  their  J=1--0
transition in Arp\,220 (consistent with both lines being optically thick). For
a comparative analysis  of HCN versus HNC excitation  using several rotational
transitions, and the diagnostic nature  of the $\rm [HNC/HCN]$ abundance ratio
for the excitation environment of  ULIRGs the interested reader is referred to
recent work by Aalto et al. (2007).

\subsection{Sub-thermal HCO$^+$ line excitation: HCO$^+$ as dense gas mass tracer}\label{section:subthermal-hco}
In both  galaxies the  low HCO$^+$  line ratios result  in LVG  solutions with
densities  mostly  $\sim 10-100$  times  lower  than  those deduced  from  the
analysis  of the  HCN ratios  (Table \ref{table:solutions}).   Thus if
only HCO$^+$  line observations were used,  they would imply a  bulk dense gas
phase  dominated  by  less  extreme  densities  than  is  actually  the  case,
especially in Arp\,220.  The several HCO$^+$ and HCN transitions observed, and
the independent analysis  afforded by the CS transitions  (for Arp\,220), make
it unlikely that the aforementioned  conclusions can be attributed to enhanced
HCN abundances (and thus thermalization  of its transitions at lower densities
because of radiative trapping) caused by  XDRs, or simply to IR-pumping of HCN
J=1--0.  Thus recent  claims that HCO$^+$ is a better tracer  of the dense gas
in (U)LIRGs than HCN (Graci\'{a}-Carpio et  al.\ 2006), do not seem to hold in
the  cases of Arp\,220  and NGC\,6240,  and a  more comprehensive  approach is
needed to decide the case (Papadopoulos 2007b).

\subsection{The mass of the dense gas phase in Arp\,220 and NGC\,6240}\label{section:dense-gas-mass}
In this section  we make use of  the constraints on the dense  gas in Arp\,220
and  NGC\,6240 obtained in  \S \ref{section:dense-phase}  to derive  the total
mass of the  dense gas in these  two systems.  We do this  using two different
methods,    the    results    of     which    are    summarized    in    Table
\ref{table:dense-gas-masses}.  The  first one assumes the  dense gas reservoir
to be fully reducible to an ensemble of self-gravitating units whose molecular
line emission does not suffer  any serious cloud-cloud shielding.  Then it can
be shown (e.g.\ Dickman,  Snell, \& Schloerb 1986) that the total  mass of e.g. 
an HCN-luminous dense gas phase can be found from

\begin{equation}
M_{\mbox{\tiny{dense}}}(\mbox{H}_2) = \alpha_{\mbox{\tiny{HCN}}} L'_{\mbox{\tiny{HCN}}},
\label{equation:method-1}
\end{equation}

\noindent
where             $\alpha_{\mbox{\tiny{HCN}}}            \simeq            2.1
\sqrt{n(\mbox{H}_2)}/T_{b,\mbox{\tiny{HCN(1-0)}}}$  is   a  conversion  factor
(e.g.\ Radford, Solomon  \& Downes 1991a).  We  insert the range of $n(\mbox{H}_2)$ and
$T_{b,\mbox{\tiny{HCN(1-0)}}}$ values inferred from the LVG solutions to the HCN line
ratios, which for Arp\,220 are $n(\mbox{H}_2)=0.3\times 10^6\,\rm{cm}^{-3}$ and
$T_{b,\rm{HCN(1-0)}} = 34-62\,$K, and for NGC\,6240 
$n(\rm{H}_2)=(0.1-0.3)\times 10^6\,\rm{cm}^{-3}$ and $T_{b,\rm{HCN(1-0)}} = 31-40\,$K.
From these values we find:          $\alpha_{\mbox{\tiny{HCN}}}\simeq
(19-34)\,\Msolar\,(\mbox{K\,km\,s$^{-1}$\,pc$^2$})^{-1}$                    and
$\alpha_{\mbox{\tiny{HCN}}}\simeq
(17-37)\,\Msolar\,(\mbox{K\,km\,s$^{-1}$\,pc$^2$})^{-1}$   for   Arp\,220  and
NGC\,6240,  which   in  turn  yields  dense   gas  mass  traced   by  HCN  of
$M_{\mbox{\tiny{dense}}}(\mbox{H}_2)  \simeq (1.8-4.2)\times 10^{10}\,\Msolar$
and   $\simeq  (1.0-2.8)\times   10^{10}\,\Msolar$  for   these   two  systems
respectively.

In a  similar fashion, the analysis of  HCO$^+$ and CS line  ratios with their
corresponding $T_b$ and $n(\mbox{H}_2)$ values yields corresponding conversion
factors  which, along  with  the luminosities  of  the lowest  HCO$^+$ and  CS
transitions  available,  yield the  dense  gas  mass  traced by  each  species
(Table~\ref{table:dense-gas-masses}).    The  constraints  of   an  underlying
density-size cloud  hierarchy (see  section 4.5) demand  $ M(\mbox{H}_2)_{\mbox{\tiny{HCN/CS}}}\leq
M_{\mbox{\tiny{HCO$^+$}}}$ (since  the HCN, CS  lines are tracing  a denser gas than  those of
HCO$^+$).   Thus  from  Table  8  we  obtain  $  M_{dense}(\mbox{H}_2)\sim  1.5\times
10^{10}\,M_{\odot}$   (Arp\,220)    and   $   M_{dense}(\mbox{H}_2)\sim   (1-2)\times
10^{10}\,M_{\odot}$  (NGC\,6240)   as  the  best  dense   gas  mass  estimates
conforming  to  the  aforementioned  inequality.   Systematic  biases  can  be
introduced by a)  the lack of a pressure-term  in eq.\ \ref{equation:method-1}
correcting  for molecular  gas  overlying the  dense  gas regions,  b) by  the
existence of  substantial stellar  mass even within  the dense gas  phase.  In
both of these cases the true conversion factor and the deduced gas masses will
be  lower  (see Bryant  \&  Scoville  1996; Downes  \&  Solomon  1998 for  the
appropriate formalism).

Finally, given  the significant  number of HCN  transition available  for both
systems, we can  estimate the dense gas  mass by assuming the bulk  of the HCN
molecules are in rotational states $J\le 6$. Then
\begin{equation}
M_{\mbox{\tiny{dense}}}(\mbox{H}_2) = [\mbox{H}_2/\mbox{HCN}] \left [ N_1 + N_2 + N_3 + N_4 + N_5 + N_6\right ] \mu m_{\mbox{\tiny{H$_2$}}},
\label{equation:method-3a}
\end{equation}
where $N_J$ is the total number of HCN molecules in state $J$ and $\mu=1.36$ accounts for He.
The HCN population numbers are calculated from 
\begin{equation}
N_J = \frac{8\pi k \nu^2_{J,J-1}}{h c^3 A_{J,J-1}} \beta^{-1}_{J,J-1} L'_{\mbox{\tiny{HCN$(J,J-1)$}}},
\label{equation:method-3b}
\end{equation}

\noindent
where $A_{J,J-1}$ and $\beta_{J,J-1} = (1-e^{-\tau_{J,J-1}})/\tau_{J,J-1}$ are
the Einstein  coefficient and  the escape probability  for the  $J \rightarrow
J-1$ transition,  respectively. Unlike the first method  this approach depends
directly  one the  assumed  $\rm [HCN/H_2]$  abundance  ratio while  it is  in
principle independent of  any assumptions regarding the velocity  field of the
dense  gas phase  (i.e. it  remains valid  for a  non-virialized,  unbound gas
phase), as long  as the HCN line emission  remains {\it effectively} optically
thin (i.e. any large $\tau $'s  emerge within small gas ``cells'').  The dense
gas masses  estimated with the  second method (see  Table 8) bracket  the best
estimates yielded by the first  one (constrained by the density-mass hierarchy
assumtion), though more  tightly for Arp\,220 than NGC\,6240. 

The  two  methods  overlap  over   the  same  range  of  $  M(\mbox{H}_2)_{dense}\sim
(1-2)\times 10^{10}\,M_{\odot}$ for both  galaxies, which is comparable to the
{\it total}  molecular gas mass  estimated independently from models  of their
interferometrically imaged CO  emission (Downes \& Solomon 1998),  or their CI
J=1--0  fluxes  (Papadopoulos \&  Greve  2004).  {\it  Thus  the  bulk of  the
  molecular ISM in these two extreme  starbursts is in a dense state with $\it
  n(H_2)\sim (10^5-10^{6})\,cm^{-3}$}.

It is  worth comparing  this mass to  the fundamental  upper limit set  by the
dynamical mass in these objects. For  Arp\,220, the latter was estimated to be
$M_{dyn}\simeq 4\times 10^{10}\,\Msolar$ (adopted  to the cosmology used here)
within a radius  of 1.4\,kpc, based on resolved observations  of CO (Downes \&
Solomon 1998).  For NGC\,6240 $M_{dyn}\simeq (0.7-1.6)\times 10^{10}\,\Msolar$
within a  radius of $\sim 500$\,pc  (Tacconi et al.\ 1999;  Bryant \& Scoville
1999).  The CO emitting region  in NGC\,6240 extends significantly beyond that
(Tacconi et  al.\ 1999)  and thus  a significant amount  of molecular  gas may
reside  beyond the  inner 500\,pc.   Assuming  a constant  rotation curve,  we
obtain  an  upper  limit   of  $M_{dyn}  =  6\times  10^{10}\,\Msolar$  within
$R=1.4\,$kpc  for NGC\,6240. Since  any HCN/CS/HCO$^+$  bright region  will be
contained  within  the CO-bright  regions  we  expect $M_{dyn}\geq  M_{\mbox{\tiny{CO}}}\geq
M_{dense}$ in both systems.  Thus the range of $M(\mbox{H}_2)_{dense}\sim (1-2)\times
10^{10}\,M_{\odot}$ corresponds  to $\sim 20\%-50\%$ of the  dynamical mass in
Arp\,220, and $\ga 15\%$ up to 100\% for NGC\,6240.

\subsection{The density-size relation for the dense gas: steeper than Larson's
law?}

In this  section we attempt  to go  a step beyond  using HCO$^+$, HCN,  and CS
lines as isolated estimators of the  dense gas mass by assuming their emission
emerging  from  an  underlying  density-size  hierarchy,  found  for  Galactic
molecular clouds by Larson  (1981), and verified by subsequent high-resolution
multi-transition studies.   The origin of the  density-size and linewidth-size
power laws revealed  for GMCs in the Galaxy may lie  in the characteristics of
supersonic turbulence and its global  driving mechanisms (Heyer \& Brunt 2004)
when present on self-gravitating structures.   As such they may remain similar
even in the extreme ISM conditions present in ULIRGs.

A  rigorous  approach  to  test  this  in  extragalactic  environments  (where
molecular clouds  cannot be viewed  with the same  level of detail)  using the
global molecular line emission  of various species would require incorporating
such power laws in the radiative transfer model used to interpret the emergent
molecular line  emission. In our case we  opt for a simpler  approach in which
the denser HCN-bright  gas is simply assumed to be  ``nested'' within the less
dense regions traced by the HCO$^+$ transitions.

We  can rudimentary  test  for  this by  deriving  the velocity-averaged  area
filling factor  of the  HCN-bright relative  to the HCO$^+$-bright gas phase:
$f_{\mbox{\tiny{HCN,HCO$^+$}}}$,  which ought to  be $<1$.  This can  be found
from
\begin{equation}
\frac{L'_{\mbox{\tiny{HCN}}}}{L'_{\mbox{\tiny{HCO$^+$}}}}=
f _{\mbox{\tiny{HCN,HCO$^+$}}} \frac{T_{b,\mbox{\tiny{HCN}}}}{T_{b,\mbox{\tiny{HCO$^+$}}}}.
\end{equation}
For the  LVG solution ranges  above $\rm T_k=40\,K$  (Table 7) and  the J=1--0
transition  of both species,  the observed  line luminosities  and LVG-deduced
intrinsic       brightness       temperatures       yield:      $0.2       \ls
f_{\mbox{\tiny{HCN,HCO$^+$}}}    \ls   1.0$    (Arp\,220)    and   $0.1    \ls
f_{\mbox{\tiny{HCN,HCO$^+$}}}  \ls   0.3$  (NGC\,6240).   For   an  underlying
density-size power  law $\rm \langle  n(H_2)\rangle \propto R^{-\alpha}  $, we
set                      $\alpha                      =                      2
log\left(n_{\mbox{\tiny{HCO$^+$}}}/n_{\mbox{\tiny{HCN}}}\right)/logf_{\mbox{\tiny{HCN,HCO$^+$}}}$
as an  approximation ($n_{\mbox{\tiny{HCN}}}$, $n_{\mbox{\tiny{HCO$^+$}}}$ are
the mean  volume densities deduced  from the LVG  fits of the HCN  and HCO$^+$
line ratios). For the range of  densities corresponding to LVG fits of the HCN
and HCO$^+$  ratios (for $T_k\geq  40\,K$), we find $\alpha\ga  2$, steeper
than  the value  of $\alpha  \sim 1$  typical for  Galactic molecular  clouds. 
Using the  HCO$^+$ and HCN J=4--3  transitions instead yields  $0.09 \ls
f_{\mbox{\tiny{HCN,HCO$^+$}}}    \ls    0.6$    (Arp\,220),   and    $0.03    \ls
f_{\mbox{\tiny{HCN,HCO$^+$}}}  \ls 0.2$  (NGC6240).  These  smaller  values are
expected for emission from the  two lines bracketing the entire range critical
densities in our line inventory  (Table 6), and an underlying density-size ISM
hierarchical    structure.     However,    even    for   these    values    of
$f_{\mbox{\tiny{HCN,HCO$^+$}}}$ the corresponding density ranges yield $\alpha
\ga 1.3$, still steeper than a  Larson-type power law.  Such a steepening could
in principle be the result of extreme tidal stripping of GMCs and then merging
of  the  surviving  densest  clumps  to super-dense  structures,  yet  another
signature  of vastly  different average  molecular gas  properties  in extreme
starbursts than those in more quiescent environments.

\subsection{The diffuse gas phase}
The observed  $^{12}$CO line  emission will have  contributions from  both the
dense  and  diffuse gas  and  the  observed global  line  ratio  of any  given
$J+1\longrightarrow J$  line with respect  to $^{12}$CO 1--0 can  therefore be
written as
\begin{equation}
r_{J+1,J} = \frac{r^{(A)}_{J+1,J} + C^{(12)}_{10} r^{(B)}_{J+1,J}}{1 + C^{(12)}_{10}},
\label{equation:line-ratio}
\end{equation}
where $r^{(A)}_{J+1,J}$  is the line ratio  for the diffuse  phase (denoted by
A), $r^{(B)}_{J+1,J}$  the one  for the  dense phase (denoted  by B),  and the
contrast factor,  $C^{(12)}_{10}$ is  discussed below.  It  is seen  from eq.\ 
\ref{equation:line-ratio}  that in order  to constrain  the properties  of the
diffuse gas, one has to determine and subtract the contribution from the dense
gas phase.  Fortunately, this can now  be done with the constraints put on the
dense phase properties in  \S \ref{section:dense-phase}.  The contrast factor,
which determines the relative emission  contribution of the two gas phases, is
given  by  $C^{(12)}_{10}  =  f_{BA}  (T^{(B)}_{b,10}/T^{(A)}_{b,10})$,  where
$f_{BA}$ is  the velocity-integrated relative  filling factor between  the two
phases  and  $T^{A,B}_{b,10}$  are  the  velocity-  and  area-integrated  line
brightness  temperatures  of  $^{12}$CO  1--0.   In order  to  determine  this
quantity, we make the reasonable assumption that the HCN emission is dominated
by the dense  gas phase and, as  a result, we can write  the global (observed)
HCN(1--0)/$^{12}$CO(1--0) line ratio as
\begin{equation}
\frac{T_{b,\mbox{\tiny{HCN}}}}{T_{b,\mbox{\tiny{CO}}}} =
 \frac{C^{(12)}_{10}}{1+{C^{(12)}_{10}}} \frac{T^{(B)}_{b,\mbox{\tiny{HCN}}}}{T^{(B)}_{b,\mbox{\tiny{CO}}}}. 
\end{equation}
We   then    estimate   $C^{(12)}_{10}$   using   the    observed   value   of
$T_{b,\mbox{\tiny{HCN}}}/T_{b,\mbox{\tiny{CO}}}$      and     the     $T^{(B)}
_{b,\mbox{\tiny{HCN}}}/T^{(B)} _{b,\mbox{\tiny{CO}}}$ values obtained from the
LVG fits of the dense gas phase. The observed global line ratio for Arp\,220 is 
$T_{b,\mbox{\tiny{HCN}}}/T_{b,\mbox{\tiny{CO}}} = 0.18\pm 0.03$, and 
$T^{(B)}_{b,\mbox{\tiny{HCN}}}/T^{(B)} _{b,\mbox{\tiny{CO}}} = 0.5-0.8$ as
estimated from the dense gas solution range. In the case of NGC\,6240 the same
two line ratios are $0.08\pm 0.01$ and $0.4-0.8$. Folding in the uncertainties on the observed
global ratios and  allowing for the range in the ratio  predicted for the dense
gas, we  find $C^{(12)}_{10} \sim 1.5-3.0$ (Arp\,220)  and $C^{(12)}_{10} \sim
1.3-2.8$ (NGC\,6240).

With $C^{(12)}_{10}$ determined, we  can then disentangle the contributions to
the  $^{12}$CO  line ratios  from  the diffuse  and  dense  phases using  Eqs\ 
\ref{equation:line-ratio}.  If we assume the $\mathcal{R}$ ratios dominated by
the diffuse phase, the resulting  $^{12}$CO and $^{13}$CO emission line ratios
for   Arp\,220   are  consistent   with   $n(\mbox{H}_2)  \simeq   (1-3)\times
10^2$\,cm\,$^{-3}$,  $T_k\ge  40\,$K,  and $\Lambda_{\mbox{\tiny{CO}}}  \simeq
(3-10)\times  10^{-5}\,(\mbox{km\,s$^{-1}$\,pc$^{-1}$})^{-1}$.  For NGC\,6240,
the   LVG   model  converges   on   $n(\mbox{H}_2)   \simeq   (1  -   3)\times
10^3$\,cm\,$^{-3}$  $T_k\ge  40\,$K,  and  $\Lambda_{\mbox{\tiny{CO}}}  \simeq
(1-3)\times  10^{-6}\,(\mbox{km\,s$^{-1}$\,pc$^{-1}$})^{-1}$.   All  solutions
correspond  to  non-virialized  $K_{vir}\gs  5$ values  (assuming  a  Galactic
abundance  [CO/H$_2$]$=10^{-4}$) and  have moderate  $^{12}$CO  J=1--0 optical
depths ($\tau  _{10}\sim 1$). This seems  to be the typical  diffuse gas phase
found  in such  systems (e.g.\ Aalto et  al.  1995;  Downes \&  Solomon 1998)
which, while dominating their low-J CO line emission, it does not contain much
of their total molecular gas mass.

\section{Dense gas mass, star formation, and its efficiency in Arp\,220 and NGC\,6240}\label{section:discussion}
The   dense  gas   masses   derived   for  Arp\,220   and   NGC\,6240  in   \S
\ref{section:dense-gas-mass},  allow  us  to  derive accurate  star  formation
efficiencies  in these two  galaxies. Although  the star  formation efficiency
(SFE) was originally defined as $\mbox{SFE} = \mbox{SFR}/M(\mbox{H}_2)$, where
SFR is the star formation rate  and $M(\mbox{H}_2)$ is the total molecular gas
mass,  we shall  here  focus on  the  SFE per  dense gas  mass,  with the  SFR
parameterized    by   the   IR    luminosity   (i.e.     $\mbox{SFE}   \propto
L_{\mbox{\tiny{IR}}}/M_{dense}$).

Using the original definition of IR  luminosity as the SED integrated from 40-
to 120-$\mu$m (Helou et  al.\ 1986), we derive $L_{\mbox{\tiny{IR}}}=1.6\times
10^{12}\,\Lsolar$      (Arp\,220)      and     $L_{\mbox{\tiny{IR}}}=7.0\times
10^{11}\,\Lsolar$  (NGC\,6240).   For the  deduced  dense  gas  mass range  of
$M_{dense}(\mbox{H}_2)\sim     (1-2)\times      10^{10}\,M_{\odot}$     we     obtain
$L_{\mbox{\tiny{IR}}}/M_{dense}\sim            (80-160)\,\Lsolar\,\Msolar^{-1}$
(Arp\,220)           and           $L_{\mbox{\tiny{IR}}}/M_{dense}           =
(35-70)\,\Lsolar\,\Msolar^{-1}$ (NGC\,6240), suggesting a somewhat higher star
formation efficiency in  Arp\,220.  Finally these values are  within the range
of those found  for a large sample of infra-red  galaxies, consisting of large
spiral galaxies, LIRGs, and a few  ULIRG by Gao \& Solomon (2004a,b) using the
HCN 1--0 line as a linear tracer of dense gas.

A  density-size hierarchy  found for  Galactic GMCs  remaining valid  in other
galaxies would suggest that HCO$^+$, HCN, and CS lines, given their increasing
excitation requirements,  would probe progressively  deeper into the  warm and
dense gas phase fueling star  formation.  We briefly explore this scenario by
calculating      three     separate      star      formation     efficiencies,
$\mbox{SFE}_{\mbox{\tiny{HCO$^+$}}}$,   $\mbox{SFE}_{\mbox{\tiny{HCN}}}$,  and
$\mbox{SFE}_{\mbox{\tiny{CS}}}$, using  the separate dense  gas mass estimates
based on HCO$^+$, HCN, and CS available for Arp\,220 only.  For this galaxy we
find $\mbox{SFE}_{\mbox{\tiny{CS}}}$ ($\simeq 107-533\,\Lsolar\,\Msolar^{-1}$)
is       larger      than       $\mbox{SFE}_{\mbox{\tiny{HCN}}}$      ($\simeq
40-94\,\Lsolar\,\Msolar^{-1}$),  which  may suggest  that  the star  formation
increases as we  probe higher densities (Shirley et al.\  2003).  On the other
hand,   this    trend   is   not   confirmed   by    HCO$^+$,   which   yields
$\mbox{SFE}_{\mbox{\tiny{HCO$^+$}}}\simeq      100-320\,\Lsolar\,\Msolar^{-1}$,
just as we do not see a significantly narrower range in $\mbox{SFE}$-values at
higher densities  as one  might otherwise expect  (Shirley et al.\  2003).  In
NGC\,6240    the   star    formation   efficiency    derived    from   HCO$^+$
($\mbox{SFE}_{\mbox{\tiny{HCO$^+$}}}\simeq  32-54\,\Lsolar\,\Msolar^{-1}$)  is
smaller       than      the       efficiency      obtained       from      HCN
($\mbox{SFE}_{\mbox{\tiny{HCN}}}\simeq   41-70\,\Lsolar\,\Msolar^{-1}$).    We
conclude that given the uncertainties in the mass estimates we cannot robustly
claim a significant increase in the SFEs derived using HCO$^+$, HCN and CS.

Theory  currently provides  some  hints that  support  a small  range of  star
formation efficiencies per  dense gas mass.  Simulations of  star formation in
turbulent media where dense ($n(\mbox{H}_2)>10^4\,$cm$^{-3}$) gas clumps decouple from
a supersonic, turbulent cascade to form stars at a constant efficiency seem to
reproduce  well some  of the  observed characteristics  of SFR  versus  gas in
galaxies (Krumholz  \& McKee 2005).  Furthermore, Scoville  (2003) showed that
the negative feedback effect of high-mass star formation naturally leads to an
upper    bound    on    the     SFE    for    star    forming    regions    of
$500\,\Lsolar\,\Msolar^{-1}$.  Although  the exact  value of this  upper limit
may change in the future (since it depends on the poorly constrained effective
radiative opacity of the dust), the underlying physics is expected to apply to
local star forming  regions as well as those  in circumnuclear starbursts.  We
note that our derived SFEs are consistent with the current upper limit.

\section{Implications for dense star forming gas at high redshifts}

Evidence  from Galactic  observations and  recent results  from  the ULIRG/QSO
Mrk\,231 (Papadopoulos,  Isaak \&  van der  Werf 2007) point  to the  same gas
phase  responsible  for both  HCN  and high-$J$  CO  line  emission.  This  is
expected, as such a phase is intimately linked with ongoing star formation and
thus its warm and dense conditions  will excite the high-$J$ CO transition and
those  of HCN, CS  and HCO$^+$  alike. Thus  the weaker  lines of  heavy rotor
molecules with frequencies within the numerous transparent atmospheric windows
at $\rm \la 350\,GHz$, can be used to probe the gas emitting also the luminous
CO  $J+1\rightarrow  J,  J+1\geq 4$  lines  at  frequencies  of $\rm  \nu  \ga
460\,GHz$  beyond which  the atmospheric  transparency window  rapidly closes,
even in sites where prime sub-mm telescopes are located.

\begin{figure}
\plotone{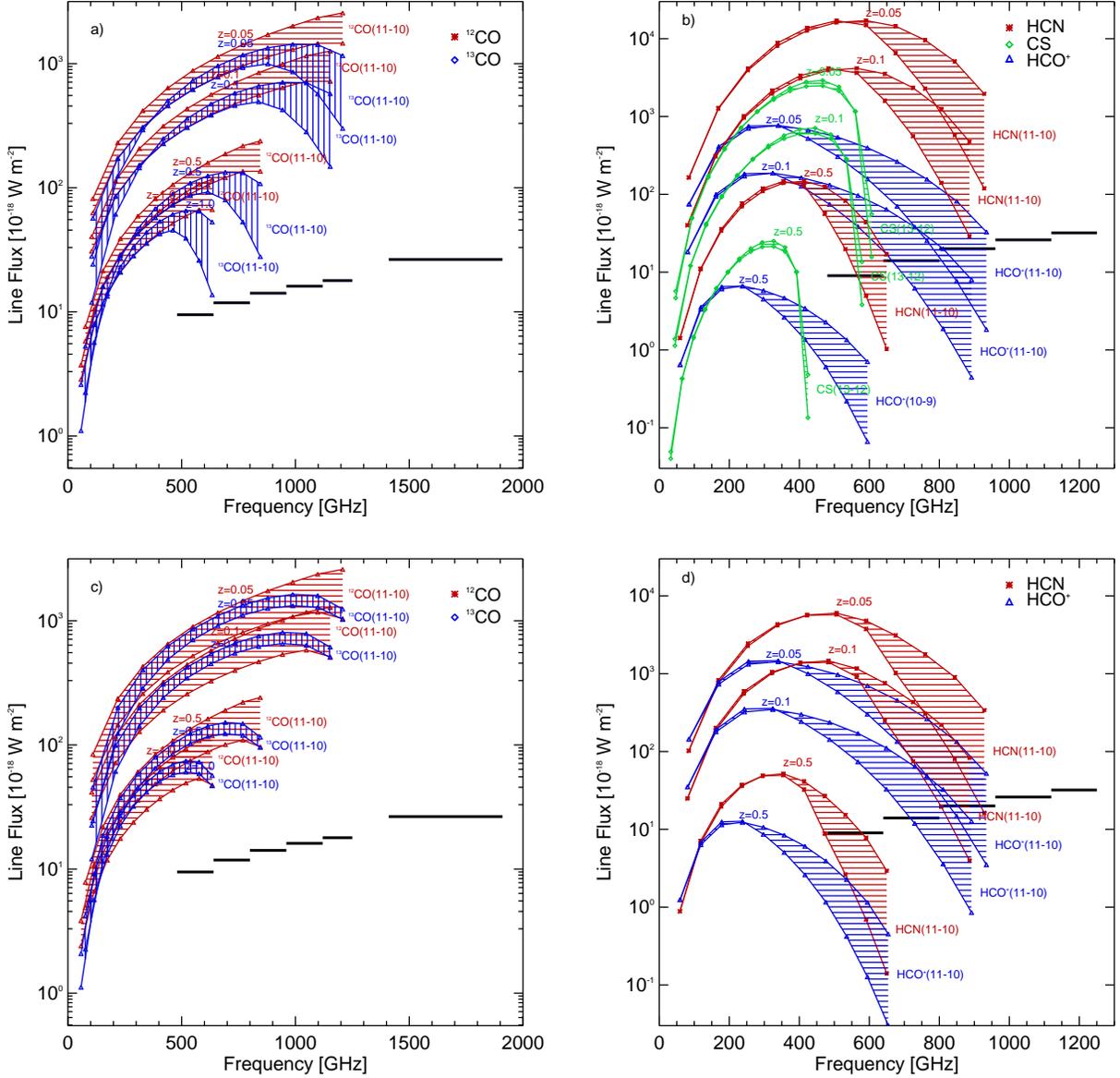} 
\caption{{\bf Top panels:} integrated line flux strengths of the full
$^{12}$CO/$^{13}$CO (left) and HCN/CS/HCO$^+$ (right) rotational ladders
predicted by the dense gas template of Arp\,220 as given by the solution ranges
in Table \ref{table:solutions}. The line fluxes are shown for Arp\,220-like
systems at redshifts $z=0.05, 0.1$, and $0.5$.  In the case of $^{12}$CO and
$^{13}$CO we have also shown the $z=1.0$ template.  {\bf Bottom panels:} Same as
above, except the constraints on the dense gas in NGC\,6240 have been used as a
template. Horizontal bars are the expected $5$-$\sigma$ sensitivity limits of
the five HI-FI bands after 1\,hr of integration.  In panels a) and c), the red
and blue curves correspond to the line templates for $^{12}$CO and $^{13}$CO,
respectively, while in panels b) and d), they correspond to HCN and HCO$^+$,
respectively. The green curve in panel b) correspond to the line template for
CS.  } \label{figure:herschel-predictions} \end{figure}

In  Fig.\ \ref{figure:herschel-predictions}(a)  and  (c) we  have plotted  the
expected line fluxes of the $^{12}$CO and $^{13}$CO line templates -- based on
Arp\,220 and NGC\,6240, respectively --  for redshifts $z=0.05, 0.1, 0.5$, and $1.0$.
The $^{12}$CO and $^{13}$CO line fluxes were extracted from the LVG model with
input conditions set by the HCN solution ranges for the dense gas in these two systems
(Table \ref{table:solutions}).
Similarly, Fig.\  \ref{figure:herschel-predictions}(b) and (d) show the
HCN/CS/HCO$^+$  line  templates for the solution ranges given in Table
\ref{table:solutions}.  For  comparison  we  also  show  the  latest
sensitivity estimates for the HI-FI  instrument (de Graauw et al.\ 1998, 2005)
-- a  series  of highly  advanced  heterodyne  receivers  due to  fly  onboard
Herschel, an  ESA cornerstone mission expected  to launch in  2008. HI-FI will
observe in six bands covering  the frequency range 480--1910\,GHz, enabling it
to target the highest $J$-transitions  in low-$z$ sources.  Fig.\ \ref{figure:herschel-predictions}
shows that if Arp\,220  and NGC\,6240 are not significantly  different from the bulk
population of  (U)LIRGs in the redshift  range $0\le z  \le 1$, HI-FI/Herschel
will be able to detect the CO $J\rightarrow J-1, J\ge 5$ lines in such objects
out to  $z\sim 0.1$ in less than  1\,hr.  At $z\sim 0.5$  the CO $J\rightarrow
J-1, J\ge 7$  lines are still within the HI-FI frequency  range, and are easily
detectable.  We see  the HCN and HCO$^+$ $J\rightarrow J-1,  J\ge 6$ lines are
detectable in systems  akin to Arp\,220 and NGC\,6240 out  to $z\sim 0.1$.  At
$z\sim 0.5$,  however, HCN  lines are still  detectable only  in Arp\,220-like
systems.  {\it The predicted HCO$^+$ lines  fluxes for both type of systems at
  $z\sim 0.5$ fall below the 1\,hr $5$-$\sigma$ sensitivity limits of HI-FI}.

Thus for (U)LIRGs in the redshift range $0\le z\ls 0.5$, HI-FI will complement
ground-based (sub)mm  facilities where only  the low-$J$ lines  are accessible
and thereby  densely sample  the rotational ladder  for a number  of important
molecules in such  systems. This will allow a full  inventory of the molecular
ISM in luminous starbursts and (U)LIRGs  and provide a valuable set of low-$z$
templates  for  interpretation of  high-$J$  detections  in extremely  distant
galaxies ($z\gs  2$) with ALMA.  Characterizing these  transitions relative to
the lower  ones in the  local extreme starburst  population -- similar  to the
need to understand local dust  SEDs before interpreting the very few frequency
points detected for  the dust continuum in high-$z$ galaxies  -- will prove to
be extremely valuable before embarking on the high-$J$, high-$z$ ventures with
ALMA.

\section{Summary}
The main results presented in this paper are:\\

\indent 1.  Using the JCMT and IRAM 30-m telescopes we report first detections
of  a   number  of  important   molecular  lines,  including   transitions  of
$^{13}\mbox{CO}$,  HCN,  CS,  and  HCO$^+$,  in the  two  prototypical  ULIRGs
Arp\,220 and  NGC\,6240.  These observations along with  measurements from the
literature complete what  is currently the largest molecular  line catalog for
such systems, capable  of probing a wide range  of densities and temperatures,
especially for their important dense and warm star-forming gas phase.

\indent  2.   In  Arp\,220 a  systematic  change  in  line profile  as  higher
densities are probed  suggests large scale gas phase  segregations within this
ULIRG.  This is not observed in NGC\,6240 even in the profiles of lines widely
separated in their  excitation requirements such as $  ^{12}$CO J=2--1 and HCN
J=4--3.  This points to the exciting possibility of using high-$J$/high-dipole
molecular lines  to deduce  the state  of gas relaxation  in ULIRGs  by simply
comparing the profiles of lines with widely different excitation requirements.
This could prove important since currently only in the closest ULIRGs can
interferometry be relied upon to reveal the state of the gas directly.\\

\indent 3.   A single  molecular gas  phase is inadequate  to account  for the
observed properties of the molecular  ISM in Arp\,220 and NGC\,6240.  Analysis
of  the numerous  line ratios  using  a two-phase  LVG model  finds a  diffuse
($n(\mbox{H}_2)\sim  10^{2-3}\,\mbox{cm}^{-3}$)  phase  that  is  most  likely
unbound, and a  dense gas phase ($n(\mbox{H}_2)\sim 10^{5-6}\,\mbox{cm}^{-3}$)
that contains the bulk of the molecular gas in these systems
($M(\mbox{H}_2)_{dense}\sim (1-2)\times 10^{10}\,M_{\odot}$).\\

\indent 4.  The  four highly excited HCN rotational  transitions (also four CS
transitions  for Arp\,220) correspond  to a  gas phase  that is  $\sim 10-100$
times denser than that suggested  by the low, sub-thermal, HCO$^+$ line ratios
measured in  both systems.  {\it  Thus HCN trace  a denser gas phase  than the
  HCO$^+$ lines.}   Moreover, these  two phases are  found consistent  with an
underlying density-size  power law, albeit  with a steeper exponent  than that
found in Galactic  GMCs (i.e. the Larson's law for  densities).  We argue that
this could be another result of the extreme tidal stripping of GMCs in
these strongly interacting systems. We also find that for CS all realistic
solutions imply $K_{vir} << 1$, which suggest that {\it the true CS abundance
is significantly higher (by an order of magnitude) than the assumed abundance
($[\mbox{CS}/\mbox{H}_2] = 10^{-9}$).} \\

\indent 5.  Finally,  using Arp\,220 and NGC\,6240 as  templates for starburst
galaxies  at higher  redshifts,  we  show that  the  HI-FI instrument  onboard
Herschel   will   be  ideal   for   detect   very   high-$J$  transitions   of
CO/HCN/CS/HCO$^+$ in such objects out  to redshifts of $z\sim 0.5$ while still
preserving  the  all  important  information  about  their  relative  velocity
profiles. Thus the  Herschel Space Observatory, in conjuction  with an ongoing
large mm/sub-mm molecular  line survey of extreme starbursts,  will be able to
produce the first  densely sampled rotational ladders of  CO, HCN, and HCO$^+$
for  (U)LIRGs in  the local  Universe,  a prerequisite  in understanding  star
formation and  its relation to dense  gas in extreme starbursts  localy and in
their high redshift counterparts.

\acknowledgments We thank the anonymous referee for a thorough report which helped
improve the paper substantially. We  are grateful to all  the telescope operators  at the IRAM
30-m and the long-standing support we enjoy from the JCMT support personel and
telescope operators.  We  are particular thankful to Jim Hoge  at the JCMT for
his  active  role in  getting  the  observations  done, often  during  adverse
technical conditions.

\clearpage



\clearpage 

%

\clearpage

\clearpage






\begin{deluxetable}{lccccccccc}
\tabletypesize{\tiny}
\tablecaption{Observational parameters for the lines observed with the JCMT. }
\tablewidth{0pt}
\tablehead{
\colhead{}                                                           & $^{12}$CO(2--1) & $^{12}$CO(3--2)  & $^{13}$CO(2--1) & $^{13}$CO(3--2)  & HCN(3--2)        & HCN(4--3)        & HCO$^+$(4--3)  & CS(5--4)         & CS(7--6)         \\
} 
\startdata
Arp\,220                                                             &                 &                  &                 &                   &                 &                  &                &                  &                  \\
      \hspace*{0.1cm}$\nu_{obs}$ [GHz]                               &   226.439       &   339.648        & 216.481         &  324.711          &  261.159        &  348.203         & 350.392        &  240.581         & 336.787          \\
      \hspace*{0.1cm}$T_{sys}$ [K]                                   &   $\sim$430     &   350-500        & 390-580         &  1020-1980        &  500-600        &  250-360         & 260-270        &  280-340         & 300-330          \\
      \hspace*{0.136cm}$t_{int}$ [s]                                 &   600           &   600            & 10800           &  2400             &  6000           &  5400            & 4800           &  5400            & 4200             \\
NGC\,6240                                                            &                 &                  &                 &                   &                 &                  &                &                  &                  \\
      \hspace*{0.1cm}$\nu_{obs}$ [GHz]                               &   225.025       &   337.527        & 215.128         &  322.682          &  259.528        &  346.027         & 348.203        &  \nodata         & 334.683          \\
      \hspace*{0.1cm}$T_{sys}$ [K]                                   &   $\sim$410     &   $\sim$290      & 400-480         &  700-810          &  280-290        &  370-750         & 310-460        &  \nodata         & 280-300          \\
      \hspace*{0.136cm}$t_{int}$ [s]                                 &   600           &   600.           & 13200           &  3600             &  3000.          &  7800            & 5400           &  \nodata         & 3000             \\
{\sc hpbw} [\arcsecs]                                                &   20.8          &   14.0           & 21.3            &   14.6            &  18.4           &  13.5            & 13.5           &  20.0            & 14.0             \\
$\eta_{ap}$                                                          &   0.60(0.58)$^a$    &   0.53(0.51)$^a$     & 0.60(0.58)$^a$      &   0.53(0.51)$^a$      &  0.54(0.54)$^a$     &  0.53(0.51)$^a$      & 0.53(0.51)$^a$     &  0.56(0.57)$^a$      & 0.53(0.51)$^a$       \\
\enddata
\tablenotetext{~}{a) Standard JCMT aperture efficiencies relevant for the observing periods in question. Values in parentheses are our own measurements.}
\label{table:jcmt-observations}
\end{deluxetable}

\begin{deluxetable}{lccccccccc|ccc}
\tabletypesize{\tiny}
\tablecaption{Observational parameters for the lines observed with the IRAM 30-m. }
\tablewidth{0pt}
\tablehead{
\colhead{}                                                           & $^{13}$CO(1--0) & $^{13}$CO(2--1) &  HCN(1--0)   & CS(2--1)  & CS(3--2)  & HNC(1--0)  & C$^{18}$O(1--0) \\
} 
\startdata
Arp\,220                                                             &                 &                 &              &           &           &            &           \\
      \hspace*{0.1cm}$\nu_{obs}$ [GHz]                               &   108.242       & 216.481         & 87.055       & 96.238    & 144.356   & 89.051     & 107.830   \\
      \hspace*{0.1cm}$T_{sys}$ [K]                                   &   $\sim$430     & 225-500         & 95-125       & 200-300   & 200-300   & 200-300    & 200-300   \\
      \hspace*{0.136cm}$t_{int}$ [s]                                 &   600           & 4320            & 4320         &           &           &            &            \\
NGC\,6240                                                            &                 &                 &              &           &           &            &           \\
      \hspace*{0.1cm}$\nu_{obs}$ [GHz]                               &   \nodata       & 215.128         &  86.512      & \nodata   & \nodata   & \nodata    & \nodata   \\
      \hspace*{0.1cm}$T_{sys}$ [K]                                   &   \nodata       & 225-500         & 96-125       & \nodata   & \nodata   & \nodata    & \nodata   \\
      \hspace*{0.136cm}$t_{int}$ [s]                                 &   \nodata       & 7200            & 7200         & \nodata   & \nodata   & \nodata    & \nodata   \\
{\sc hpbw} [\arcsecs]                                                &   22.4          & 11.4            & 28.3         & 25.1      & 18.0      & 27.1       & 22.4      \\
$\eta_{ap}$                                                          &   0.56          & 0.41            & 0.62         & 0.59      & 0.52      & 0.59       & 0.56      \\
\enddata
\label{table:iram30m-observations}
\end{deluxetable}

\begin{deluxetable}{lcccll|cccll}
\rotate
\tabletypesize{\scriptsize}
\tablecaption{Molecular and atomic transitions detected towards Arp\,220 and NGC\,6240.}
\tablewidth{0pt}
\tablehead{
 & \multispan4{\hfil Arp\,220 \hfil}  & \multispan6{\hfil NGC\,6240 \hfil} \\
\colhead{Line}            & \colhead{{\sc fwhm}$^a$}  &  \colhead{$S_{\nu} \Delta v^{b)}$}   &  \colhead{$L'^{c)}$}        & \colhead{Telescope} & \colhead{Ref.}  & \colhead{{\sc fwhm}$^a$}            &  \colhead{$S_{\nu}\Delta v^{b)}$}   &  \colhead{$L'^{c)}$}      & \colhead{Telescope}  & \colhead{Ref.}}
\startdata
$^{12}$CO(1--0)           & $504$                     &  $410\pm 41$                         &  $5.9\pm 0.6$               & OVRO                &   [1]           & $370$                               &  $314\pm 64$                        &  $8.3\pm 1.7$            & IRAM 30m             &  [3]\\
                          & \nodata                   &  $384\pm 115^d$                      &  $5.5\pm 1.7$               & OVRO                &   [2]           & $469$                               &  \underline{$324\pm 33$}            &  \underline{$8.6\pm 0.9$}& OVRO                 &  [18]\\
                          & $480$                     &  \underline{$496\pm 99$}             &  \underline{$7.2\pm 1.4$}   & IRAM 30m            &   [3]           &                                     &  $322\pm 29$                        &  $8.5\pm 0.8$            &                      &  \\
                          &                           &  $419\pm 36$                         &  $6.0\pm 0.5$               &                     &                 &                                     &                                     &                          &                      &  \\
                          &                           &                                      &                             &                     &                 &                                     &                                     &                          &                      &  \\
$^{12}$CO(2--1)           & $447$                     &  $1071\pm 107$                       &  $3.9\pm 0.4$               & OVRO                &   [1]           & $300$                               &  $1220\pm 366^d$                    &  $8.1\pm 2.4$            & IRAM PdBI            &  [19]\\
                          & \nodata                   &  $1088\pm 330^d$                     &  $3.9\pm 1.2$               & IRAM 30m            &   [4]           & $410$                               &  \underline{$1741\pm 350$}          &  \underline{$11.6\pm 2.3$}   & JCMT             &  This work\\
                          & \nodata                   &  $1118\pm 112$                       &  $4.0\pm 0.4$               & NRAO 12m            &   [5]           &                                     &  $1492\pm 253$                      &  $9.9\pm 1.7$            &                      &\\
                          & $380$                     &  $1100\pm 220$                       &  $4.0\pm 0.8$               & OVRO                &   [6]           &                                     &                                     &                          &                      & \\
                          &                           &  $1730\pm 519$                       &  $6.2\pm 1.9$               & JCMT                &   [9]           &                                     &                                     &                          &                      & \\
                          & $420$                     &  \underline{$1549\pm 311$}           &  \underline{$5.6\pm 1.1$}   & JCMT                &   This work     &                                     &                                     &                          &                      & \\
                          &                           &  $1127\pm 69$                        &  $4.1\pm 0.3$               &                     &                 &                                     &                                     &                          &                      & \\
                          &                           &                                      &                             &                     &                 &                                     &                                     &                          &                      &  \\
$^{12}$CO(3--2)           & \nodata                   &  $4094\pm 758$                       &  $6.6\pm 1.2$               & JCMT                &   [7]           & $390$                               &  $3205\pm 642$                      &  $9.5\pm 1.9$            & JCMT                 &  This work \\
                          & $306$                     &  $4154\pm 975$                       &  $6.7\pm 1.6$               & HHT                 &   [8]           &                                     &                                     &                          &                      & \\
                          &                           &  $3700\pm 1110^d$                    &  $5.9\pm 1.8$               & JCMT                &   [9]           &                                     &                                     &                          &                      & \\
                          & $510$                     &  \underline{$3168\pm 634$}           &  \underline{$5.1\pm 1.0$}   & JCMT                &   This work     &                                     &                                     &                          &                      & \\
                          &                           &  $3674\pm 405$                       &  $5.9\pm 0.6$               &                     &                 &                                     &                                     &                          &                      & \\
                          &                           &                                      &                             &                     &                 &                                     &                                     &                          &                      &  \\
$^{13}$CO(1--0)           & \nodata                   &  $\le 9.4$                           &  $\le 0.15$                 & SEST                &   [5]           & \nodata                             &  $6.5\pm 1.9$                       & $0.19\pm 0.06$           & IRAM 30m             & [20] \\       
                          & $369$                     &  $9\pm 2$                            &  $0.14\pm 0.03$             & IRAM 30m            &   This work     &                                     &                                     &                          &                      &  \\       
                          &                           &                                      &                             &                     &                 &                                     &                                     &                          &                      &  \\
$^{13}$CO(2--1)           & \nodata                   &  $56\pm 17^d$                        &  $0.22\pm 0.07$             & IRAM 30m            &   [10]           & $360$                               &  $28\pm 7$                          & $0.20\pm 0.05$           & JCMT                 & This work  \\
                          & \nodata                   &  $54\pm 11$                          &  $0.21\pm 0.04$             & JCMT                &   [11]          & $400$                               &  \underline{$25\pm 5$}              & \underline{$0.18\pm 0.04$} & IRAM 30m           & This work\\
                          & $450$                     &  $60\pm 13$                          &  $0.24\pm 0.05$             & IRAM 30m            &   This work     &                                     &  $26\pm 4$                          & $0.19\pm 0.03$           &                      &\\
                          & $390$                     &  \underline{$70\pm 16$}              &  \underline{$0.28\pm 0.06$} & JCMT                &   This work     &                                     &                                     &                          &                      &\\
                          &                           &  $59\pm 7$                           &  $0.23\pm 0.03$             &                     &                 &                                     &                                     &                          &                      & \\
                          &                           &                                      &                             &                     &                 &                                     &                                     &                          &                      &  \\
$^{13}$CO(3--2)           & $410$                     &  $404\pm 86$                         &  $0.71\pm 0.15$             & JCMT                &   This work     & \nodata                             &  $\le 79$                           & $\le 0.3$                & JCMT                 & This work\\
                          &                           &                                      &                             &                     &                 &                                     &                                     &                          &                      &  \\
$^{12}$C$^{18}$O(1--0)    & $280$                     &  $9\pm 2$                            &  $0.14\pm 0.03$             & IRAM 30m            &   This work     & \nodata                             &  \nodata                            & \nodata                  & \nodata              & \nodata \\
                          &                           &                                      &                             &                     &                 &                                     &                                     &                          &                      &  \\
HNCO(5--4)                & $280$                     &  $5\pm 2$                            &  $0.08\pm 0.03$             & IRAM 30m            &   This work     & \nodata                             &  \nodata                            & \nodata                  & \nodata              & \nodata \\
                          &                           &                                      &                             &                     &                 &                                     &                                     &                          &                      &  \\
HCN(1--0)                 & \nodata                   &  $35\pm 11^d$                        &  $0.85\pm 0.27$             & IRAM PdBI           &   [12]          & \nodata                             &  $26\pm8^d$                         & $1.17\pm0.36$            & IRAM 30m             & [13]\\
                          & $550$                     &  $59\pm 12$                          &  $1.44\pm 0.29$             & IRAM 30m            &   This work     & \nodata                             &  $14\pm 4^d$                        & $0.63\pm 0.18$           & NMA                  & [21]\\
                          & \nodata                   &  $38\pm 12$                          &  $0.93\pm 0.29$             & IRAM 30m            &   [13]          & $340$                               &  $13\pm 3$                          & $0.58\pm 0.13$           & IRAM 30m            & This work\\
                          & $530$                     &  \underline{$48\pm 14$}              &  \underline{$1.17\pm 0.34$} & IRAM 30m            &   [22]          & $410$                               &  \underline{$16\pm 5$}              & \underline{$0.72\pm 0.22$}& IRAM 30m           & [22]\\
                          &                           &  $44\pm 6$                           &  $1.08\pm 0.15$             &                     &                 &                                     &  $15\pm 2$                          & $0.66\pm 0.09$           &                      & \\
                          &                           &                                      &                             &                     &                 &                                     &                                     &                          &                      &  \\
HCN(2--1)                 & $530$                     &  $144\pm 43$                         &  $0.88\pm 0.26$             & IRAM 30m            &   [22]          & $410$                               &  $45\pm 13$                         & $0.51\pm 0.15$           & IRAM 30m             & [22]\\
                          &                           &                                      &                             &                     &                 &                                     &                                     &                          &                      &  \\
HCN(3--2)                 & $540$                     &  $361\pm 73$                         &  $0.98\pm 0.20$             & JCMT                &   This work     & $430$                               &  $160\pm 33$                        & $0.80\pm 0.16$           & JCMT                 & This work\\
                          & $530$                     &  \underline{$258\pm 77$}             &  \underline{$0.70\pm 0.21$} & IRAM 30m            &   [22]          & $410$                               &  \underline{$74\pm 22$}             & \underline{$0.37\pm 0.11$}& IRAM 30m             & [22]\\
                          &                           &  $312\pm 53$                         &  $0.85\pm 0.14$             &                     &                 &                                     &  $160\pm 33$                        & $0.80\pm 0.16$           &                      &  \\
                          &                           &                                      &                             &                     &                 &                                     &                                     &                          &                      &  \\
HCN(4--3)                 & $590$                     &  $587\pm 118$                        &  $0.90\pm 0.18$             & JCMT                &   This work     & $320$                               &  $138\pm 29$                        & $0.39\pm 0.08$           & JCMT                 & This work\\
                          &                           &                                      &                             &                     &                 &                                     &                                     &                          &                      &  \\
HNC(1--0)                 & $516\pm 38$               &  $55\pm 11$                          &  $1.28\pm 0.26$             & IRAM 30m            &   [14]          & \nodata                             &  $14\pm 7$                          & $0.60\pm0.30$            & SEST                 & [15]\\
                          & \nodata                   &  $34\pm 7$                           &  $0.79\pm 0.16$             & SEST                &   [15]          &                                     &                                     &                          &                      &     \\
                          & $407$                     &  \underline{$43\pm 8$}               &  \underline{$0.96\pm 0.19$} & IRAM 30m            &   This work     &                                     &                                     &                          &                      &  \\
                          &                           &  $40\pm 5$                           &  $0.94\pm 0.11$             &                     &                 &                                     &                                     &                          &                      &  \\
                          &                           &                                      &                             &                     &                 &                                     &                                     &                          &                      &  \\
CS(2--1)                  & $339$                     &  $12\pm 3$                           &  $0.24\pm 0.06$             & IRAM 30m            &   This work     & \nodata                             &  \nodata                            & \nodata                  & \nodata              & \nodata\\
                          &                           &                                      &                             &                     &                 &                                     &                                     &                          &                      &  \\
CS(3--2)                  & \nodata                   &  $22\pm 7^d$                         &  $0.20\pm 0.06$             & IRAM 30m            &   [10]           & \nodata                             &  \nodata                            & \nodata                  & \nodata              & \nodata\\
                          & $455$                     &  \underline{$26\pm 5$}               &  \underline{$0.24\pm 0.04$} & IRAM 30m            &   This work     &                                     &                                     &                          &                      & \\
                          &                           &  $25\pm 4$                           &  $0.23\pm 0.03$             &                     &                 &                                     &                                     &                          &                      & \\
                          &                           &                                      &                             &                     &                 &                                     &                                     &                          &                      &  \\
CS(5--4)                  & $460$                     &  $68\pm 15$                          &  $0.22\pm 0.05$             & JCMT                &   This work     & \nodata                             &  \nodata                            & \nodata                  & \nodata              & \nodata\\
                          &                           &                                      &                             &                     &                 &                                     &                                     &                          &                      &  \\
CS(7--6)                  & $390$                     &  $111\pm 24$                         &  $0.18\pm 0.04$             & JCMT                &   This work     & $240$                               &  $60\pm 14$                         & $0.18\pm 0.04$           & JCMT                 & This work\\
                          &                           &                                      &                             &                     &                 &                                     &                                     &                          &                      &  \\
HCO$^+$(1--0)             & \nodata                   &  $20\pm 6^d$                         &  $0.48\pm 0.14$             & IRAM PdBI           &   [12]          & $420$                               &  $21\pm 3$                          & $0.93\pm 0.13$           & NMA                  & [21]\\
                          & \nodata                   &  $19\pm 6$                           &  $0.46\pm 0.14$             & IRAM 30m            &   [13]          & \nodata                             &  \underline{$25\pm 8^{d,e}$}        & \underline{$1.11\pm 0.35$} & IRAM 30m           & [16]\\
                          & \nodata                   &  \underline{$22\pm 7^{d,e}$}         &  \underline{$0.53\pm 0.17$} & IRAM 30m            &   [16]          &                                     &  $21\pm 3$                          & $0.95\pm 0.12$           &                      & \\
                          &                           &  $20\pm 4$                           &  $0.49\pm 0.09$             &                     &                 &                                     &                                     &                          &                      & \\
                          &                           &                                      &                             &                     &                 &                                     &                                     &                          &                      &  \\
HCO$^+$(3--2)             & \nodata                   &  $49\pm 15^{d,e}$                    &  $0.13\pm 0.04$             & IRAM 30m            &   [16]          & \nodata                             &  $47\pm 14^{d,e}$                   & $0.23\pm 0.07$           & IRAM 30m             & [16]\\
                          &                           &                                      &                             &                     &                 &                                     &                                     &                          &                      &  \\
HCO$^+$(4--3)             & $280$                     &  $106\pm 23$                         &  $0.16\pm 0.03$             & JCMT                &   This work     &  $400$                              &  $71\pm 17$                         & $0.20\pm0.05$            & JCMT                 & This work\\
                          &                           &                                      &                             &                     &                 &                                     &                                     &                          &                      &  \\
C\,{\sc i}($^3P_1-^3P_0$) &                           &  $1160\pm 350$                       &                             & JCMT                &   [17]          &                                     &  $600\pm 180$                       &                          & JCMT                 & [17]\\
\enddata
\tablenotetext{~}{a) Line widths are in km\,s$^{-1}$.\\
\hspace*{0.3cm} b) Line flux densities are in units of Jy\,km\,s$^{1}$.\\
\hspace*{0.3cm} c) Line luminosities are in units of $\times 10^9$\,K\,km\,s$^{-1}$\,pc$^2$.\\
\hspace*{0.3cm} d) We assume a 30 per cent error on the flux measurement.\\
\hspace*{0.3cm} e) The fluxes are derived from the luminosities quoted by Graci\'{a}-Carpio et al.\ (2006).}
\tablenotetext{~}{[1] Scoville et al.\ (1997); [2] Scoville et al.\ (1991); [3] Solomon et al.\ (1997); [4] Radford et al.\ (1991a); 
[5] Aalto et al.\ (1995); [6] Downes \& Solomon (1998); [7] Lisenfeld et al.\ (1996); [8] Mauersberger et al.\ (1999); [9] Wiedner et al.\ (2002); [10] Solomon et al.\ (1990);
[11] Papadopoulos \& Seaquist (1998); [12] Radford et al.\ (1991b); [13] Solomon et al.\ (1992a); [14] H\"{u}ttemeister et al.\ (1995); [15] Aalto et al.\ (2002); [16] Grac\'ia-Carpio et al.\ (2006);
[17] Papadopoulos \& Greve (2004); [18] Bryant et al.\ (1999); [19] Tacconi et al.\ (1999); [20] Casoli, Dupraz \& Combes (1992); [21] Nakanishi et al.\ (2005); [22] Krips et al.\ (2007).}
\label{table:lines}
\end{deluxetable}

\begin{deluxetable}{lcccccccccc}
\tabletypesize{\scriptsize}
\tablecaption{CO line ratios for Arp\,220 and NGC\,6240.}
\tablewidth{0pt}
\tablehead{
\colhead{Galaxy} & \colhead{$\frac{^{12}\mbox{CO(2 -- 1)}}{^{12}\mbox{CO(1 -- 0)}}$}
                 & \colhead{$\frac{^{12}\mbox{CO(3 -- 2)}}{^{12}\mbox{CO(1 -- 0)}}$}   
                 & \colhead{$\frac{^{12}\mbox{CO(1 -- 0)}}{^{13}\mbox{CO(1 -- 0)}}$}   
                 & \colhead{$\frac{^{12}\mbox{CO(2 -- 1)}}{^{13}\mbox{CO(2 -- 1)}}$}   
                 & \colhead{$\frac{^{12}\mbox{CO(3 -- 2)}}{^{13}\mbox{CO(3 -- 2)}}$} 
                 & \colhead{$\frac{\mbox{C$^{18}$O(1 -- 0)}}{^{13}\mbox{CO(1 -- 0)}}$} \\
\colhead{}       & \colhead{$r_{21}$}
                 & \colhead{$r_{32}$}   
                 & \colhead{$\mathcal{R}_{10}$}   
                 & \colhead{$\mathcal{R}_{21}$}   
                 & \colhead{$\mathcal{R}_{32}$} 
                 & \colhead{$\mathcal{R}^{(18)}_{10}$}
}
\startdata
Arp\,220         & $0.7\pm 0.1$  & $1.0\pm 0.1$  & $43\pm 10$ & $18\pm 3$   & $8\pm 2$  & $1.0\pm 0.3$\\
NGC\,6240        & $1.2\pm 0.2$  & $1.1\pm 0.2$  & $45\pm 15$ & $52\pm 12$  & $\ge 32$  & \nodata\\
\enddata
\label{table:line-ratios-diffuse}
\end{deluxetable}

\begin{deluxetable}{lccccccccc}
\tabletypesize{\tiny}
\tablecaption{Observed and modeled HCN, CS and HCO$^+$ line ratios for Arp\,220 and NGC\,6240.}
\tablewidth{0pt}
\tablehead{
\colhead{Galaxy} & \colhead{$\frac{\mbox{HCN(2 -- 1)}}{\mbox{HCN(1 -- 0)}}$}
                 & \colhead{$\frac{\mbox{HCN(3 -- 2)}}{\mbox{HCN(1 -- 0)}}$}
                 & \colhead{$\frac{\mbox{HCN(4 -- 3)}}{\mbox{HCN(1 -- 0)}}$}   
                 & \colhead{$\frac{\mbox{HCN(1 -- 0)}}{\mbox{HNC(1 -- 0)}}$}   
                 & \colhead{$\frac{\mbox{CS(3 -- 2)}}{\mbox{CS(2 -- 1)}}$}   
                 & \colhead{$\frac{\mbox{CS(5 -- 4)}}{\mbox{CS(2 -- 1)}}$}   
                 & \colhead{$\frac{\mbox{CS(7 -- 6)}}{\mbox{CS(2 -- 1)}}$} 
                 & \colhead{$\frac{\mbox{HCO$^+$(3 -- 2)}}{\mbox{HCO$^+$(1 -- 0)}}$} 
                 & \colhead{$\frac{\mbox{HCO$^+$(4 -- 3)}}{\mbox{HCO$^+$(1 -- 0)}}$} 
}
\startdata
Arp\,220         &                  &                &                 &                &              &              &              &                &                \\  
~~~observed      & $0.8\pm 0.3$     & $0.8\pm 0.2$   &  $0.8\pm 0.2$   & $1.2\pm 0.2$   & $1.0\pm 0.3$ & $1.0\pm 0.3$ & $0.8\pm 0.3$ & $0.27\pm 0.10$ & $0.33\pm 0.09$  \\
~~~modeled       & $1.0$            & $0.9$          &  $0.8$          & \nodata        & $1.0$        & $0.9-1.0$    & $0.8-0.9$    & $0.3-0.4$      & $0.4$  \\
NGC\,6240        &                  &                &                 &                &              &              &              &                &                 \\
~~~observed      & $0.8\pm 0.3$     & $1.2\pm 0.3$   &  $0.6\pm 0.1$   & $1.1\pm 0.6$   & \nodata      & \nodata      & \nodata      & $0.24\pm 0.08$ & $0.21\pm 0.06$  \\
~~~modeled       & $0.9-1.0$        & $0.8-0.9$      &  $0.6-0.7$      & \nodata        & \nodata      & \nodata      & \nodata      & $0.3-0.4$      & $0.4$  \\
\enddata
\label{table:line-ratios-dense}
\end{deluxetable}

\clearpage
\begin{deluxetable}{lccc}
\tabletypesize{\tiny}
\tablecaption{Line excitation characteristics (from Jansen 1995).}
\tablewidth{0pt}
\tablehead{
\colhead{Transition}       & $\nu_{\mbox{\tiny{rest}}}$  &   $E_u/k_B$    & $n_{\mbox{\tiny{crit}}}$ \\ 
                           & [GHz]                       &   [K]          & [cm$^{-3}$]} 
\startdata
$^{12}$CO(1--0)            & 115.271                     &   5.5          & $4.1\times 10^2$  \\ 
$^{12}$CO(2--1)            & 230.538                     &   16.6         & $2.7\times 10^3$  \\ 
$^{12}$CO(3--2)            & 345.796                     &   33.2         & $8.4\times 10^3$  \\ 
HCN(1--0)                  & 88.632                      &   4.3          & $2.3\times 10^5$  \\ 
HCN(3--2)                  & 265.886                     &   25.5         & $4.1\times 10^6$  \\ 
HCN(4--3)                  & 354.734                     &   42.5         & $8.5\times 10^6$  \\ 
HCO$^+$(1--0)              & 89.189                      &   4.3          & $3.4\times 10^4$  \\ 
HCO$^+$(3--2)              & 267.558                     &   25.7         & $7.8\times 10^5$  \\ 
HCO$^+$(4--3)              & 356.734                     &   42.8         & $1.8\times 10^6$  \\ 
CS(2--1)                   & 97.980                      &   7.1          & $8.0\times 10^4$  \\ 
CS(3--2)                   & 146.969                     &   14.1         & $2.5\times 10^5$  \\ 
CS(5--4)                   & 244.936                     &   35.3         & $1.1\times 10^6$  \\ 
CS(7--6)                   & 342.883                     &   65.8         & $2.9\times 10^6$  \\ 
\enddata
\tablenotetext{~}{$n_{\mbox{\tiny{crit}}}=A_{ul}/\sum_{i}C_{ui}$, where $C_{ui}$ is the collisional rate
'out' of the level $u$ (either up or down), calculated for $T_k=100\,$K in the optically thin limit.}
\label{table:line-data}
\end{deluxetable}

\begin{deluxetable}{llllll}
\tabletypesize{\tiny}
\tablecaption{LVG solution ranges, see \S \ref{section:dense-phase} for details. The most likely solutions
have been highlighted in bold.}
\tablewidth{0pt}
\tablehead{
                           & $T_k$/[K]                  & $n(\mbox{H}_2)$/cm$^{-3}$ &  $\Lambda$/$\rm (km\,s^{-1}\,pc^{-1})^{-1}$   & $\chi^{2 \dagger}_{\nu}$  &  $K_{vir}$
   } 
\startdata
                           \hline
Arp\,220:                  &                            &                           &                                               &                         &         \\                           
~~~~HCO$^+$                &                            &                           &                                               &                         &          \\
                           & $10 - 40$                  & $(1-3)\times 10^4$        &  $(0.3-1)\times 10^{-8}$                      &  $2.1-2.8$              & $0.4-0.7$\\
                           & $\mathbf{45 - 95}$         & $\mathbf{1\times 10^4}$   &  $\mathbf{0.3\times 10^{-8}}$                 &  $\mathbf{2.1}$         & $\mathbf{1.2}$\\
                           & $\mathbf{100 - 120}$       & $\mathbf{0.3\times 10^4}$ &  $\mathbf{1\times 10^{-8}}$                   &  $\mathbf{2.0-2.1}$     & $\mathbf{0.7}$\\
~~~~HCN                    &                            &                           &                                               &                         &          \\
                           & $10 - 25$                  & $(0.3-1)\times 10^6$      &  $(0.3-1)\times 10^{-7}$                      &  $0.5-1.3$              & $0.02-0.17$\\
                           & $30 - 40$                  & $1\times 10^6$            &  $0.3\times 10^{-9}$                          &  $0.7-0.8$              & $3.1$\\
                           & $\mathbf{45 - 120}$        & $\mathbf{0.3\times 10^6}$ &  $\mathbf{0.1\times 10^{-8}}$                 &  $\mathbf{0.7-0.8}$     & $\mathbf{1.7}$\\
~~~~CS                     &                            &                           &                                               &                         &            \\
                           & $10 - 30$                  & $10\times 10^6$           &  $(0.001-0.3)\times 10^{-8}$                  &  $0.2-0.5$              & $0.005-1.5$\\
                           & $35 - 40$                  & $3\times 10^6$            &  $0.01\times 10^{-8}$                         &  $0.2-0.3$              & $0.3$\\
                           & $\mathbf{45 - 80}$         & $\mathbf{1\times 10^6}$   &  $\mathbf{0.1\times 10^{-8}}$                 &  $\mathbf{0.2-0.3}$     & $\mathbf{0.05}$\\
                           & $\mathbf{85 - 120}$        & $\mathbf{0.3\times 10^6}$ &  $\mathbf{(0.3-1)\times 10^{-8}}$             &  $\mathbf{0.3}$         & $\mathbf{0.009-0.03}$\\
\hline
NGC\,6240:                 &                            &                           &                                               &                         &         \\                           
~~~~HCO$^+$                &                            &                           &                                               &                         &          \\
                           & $10 - 25$                  & $(1-3)\times 10^4$        &  $(0.3-1)\times 10^{-8}$                      &  $1.9-2.7$              & $0.4-0.7$\\
                           & $\mathbf{30 - 55}$         & $\mathbf{1\times 10^4}$   &  $\mathbf{0.3\times 10^{-8}}$                 &  $\mathbf{1.8-1.9}$     & $\mathbf{1.2}$\\
                           & $\mathbf{60 - 120}$        & $\mathbf{0.3\times 10^4}$ &  $\mathbf{1\times 10^{-8}}$                   &  $\mathbf{1.6-1.7}$     & $\mathbf{0.7}$\\
~~~~HCN                    &                            &                           &                                               &                         &          \\
                           & $10 - 20$                  & $(0.1-1)\times 10^6$      &  $(0.3-1)\times 10^{-7}$                      &  $1.3-1.7$              & $0.03-0.1$\\
                           & $25 - 30$                  & $0.3\times 10^6$          &  $0.1\times 10^{-8}$                          &  $1.3-1.4$              & $1.7$\\
                           & $\mathbf{35 - 55}$         & $\mathbf{0.1\times 10^6}$ &  $\mathbf{1\times 10^{-8}}$                   &  $\mathbf{1.3-1.5}$     & $\mathbf{0.3}$\\
                           & $\mathbf{60 - 120}$        & $\mathbf{0.3\times 10^6}$ &  $\mathbf{0.03\times 10^{-8}}$                &  $\mathbf{1.4-1.5}$     & $\mathbf{5.5}$\\
\enddata
\label{table:solutions}
\tablenotetext{~}{$\dagger$ the chi-squared fit was calculated as $\chi^{2}_{\nu} = \Sigma_i \frac{1}{\sigma_i}(R_{obs} - R_{model})^2$, where $R_{obs}$ and
$R_{model}$ are the observed and modeled line ratios, respectively, and $\sigma_i$ is the associated error.}
\end{deluxetable}



\begin{deluxetable}{lcc}
\tabletypesize{\tiny}
\tablecaption{Estimates of the total dense gas mass in Arp\,220 and NGC\,6240. See \S \ref{section:dense-gas-mass} for details.}
\tablewidth{0pt}
\tablehead{
 & \multispan2{\hfil$M_{dense} [\times 10^{10}\,\Msolar]$ \hfil}  } 
\startdata
                                        & Arp\,220                         & NGC\,6240\\
Method 1                                &                                  &                               \\ 
\hspace*{1cm}    HCO$^+$(1--0)          & 0.5-1.6                          & 1.3-2.2         \\
\hspace*{1cm}    HCN(1--0)              & 1.8-4.2                          & 1.0-2.8                   \\
\hspace*{1cm}    CS(2--1)               & 0.3-1.5                          & . . . \\
Method 2                                &                                  &      \\ 
\hspace*{1cm}    HCN                    & 1.4-7.0                          & 1.7-5.7\\
\hline
Best Estimate                           & 0.3-1.6                          & 1.0-2.2\\
\enddata
\label{table:dense-gas-masses}
\end{deluxetable}



\begin{thebibliography}{}
\bibitem[Aalto et al.\ 1995]{Aalto-et-al-1995} Aalto S., Booth R.\ S., Black J.\ H., Johansson L.\ E.\ B.\ 1995, A\&A, 300, 369.
\bibitem[Aalto et al.\ 2002]{Aalto-et-al-2002} Aalto S., Polatidis A.\ G., H\"{u}ttemeister S., Curran S.\ J.\ 2002, A\&A, 381. 783.
\bibitem[Aalto et al.\ 2007]{Aalto-et-al-2007} Aalto S., Spaans M., Wiedner M. C., H\"uttemeister 2007, A\&A, 464, 193
\bibitem[Bayet et al.\ 2004]{Bayet-et-al-2004} Bayet E., Gerin M., Phillips T.\ G., Contursi A.\ 2004, A\&A, 427, 45.
\bibitem[Bergin et al.\ 1996]{Bergin-et-al-1996} Bergin E.\ A., Snell R.\ L.\ \& Goldsmith P.\ F.\ 1996, ApJ, 460, 343. 
\bibitem[Bradford et al.\ 2003]{Bradford-et-al-2003} Bradford C.\ M., Nikola T., Stacey G.\ J., Bolatto A.\ D., Jackson J.\ M., Savage M.\ L., Davidson J.\ A., Higdon S.\ J.\ 2003, ApJ, 586, 891.
\bibitem[Bryant \& Scoville 1996]{Bryant-and-Scoville-1996} Bryant P.\ M.\ \& Scoville, N.\ Z.\ 1996, ApJ, 457, 678.
\bibitem[Bryant \& Scoville 1999]{Bryant-and-Scoville-1999} Bryant P.\ M.\ \& Scoville, N.\ Z.\ 1999, AJ, 117, 263.
\bibitem[Casassus, Stahl \& Wilson 2005]{Casassus-Stahl-Wilson-2005} Casassus S., Stahl O.\ \& Wilson T.\ L.\ 2005, A\&A, 441, 181.
\bibitem[Casoli et al.\ 1992]{Casoli-et-al-1992} Casoli F,. Dupraz C.\ \& Combes F.\ 1992, A\&A, 264, 55.
\bibitem[Downes \& Solomon 1998]{Downes-and-Solomon-1998} Downes D.\ \& Solomon P.\ M.\ 1998, ApJ, 507, 615.
\bibitem[Dopita-et-al-2005]{Dopita-et-al-2005} Dopita M.\ A., et al.\ 2005, ApJ, 619, 755.
\bibitem[Dickman, Snell, \& Schloerb 1986]{Dickman-Snell-Schloerb-1986} Dickman R. L., Snell R. L.\ \& Schloerb F. P. 1986, ApJ, 309, 326
\bibitem[Flower \& Launay 1985]{Flower-and-Launay-1985} Flower D.\ R.\ \& Launay J.\ M.\ 1985, MNRAS, 214, 271. 
\bibitem[Flower 1999]{Flower-1999} Flower D.\ R.\ 1999, MNRAS, 305, 651.
\bibitem[Gao and Solomon 2004a]{Gao-and-Solomon-2004a} Gao Y.\ \& Solomon P.\ M.\ 2004a, ApJS, 152, 63.
\bibitem[Gao and Solomon 2004b]{Gao-and-Solomon-2004b} Gao Y.\ \& Solomon P.\ M.\ 2004b, ApJ, 606, 271.
\bibitem[Goldsmith \& Langer 1978]{Goldsmith-and-Langer-1978} Goldsmith P.\ F.\ \& Langer W.\ D.\ 1978, ApJ, 222, 881.
\bibitem[Goldsmith et al.\ 1981]{Goldsmith-et-al-1981} Goldsmith P.\ F., Langer W.\ D., Ellder J., Kollberg E., Irvine W.\ 1981, ApJ, 249, 524.
\bibitem[Goldsmith 2001]{Goldsmith-2001} Goldsmith P.\ F.\ 2001, ApJ, 557, 736.
\bibitem[Gordon et al.\ 1992]{Gordon-et-al-1992} Gordon M.\ A., Baars J.\ W.\ M.\ \& Cocke W.\ J.\ 1992, A\&A, 264, 337.
\bibitem[Graci\'{a}-Carpio et al.\ 2006]{Gracia-Carpio-et-al-2006} Graci\'{a}-Carpio J., Garc\'{i}a-Burillo S., Planesas P., Colina L.\ 2006, ApJ, 640, L135.
\bibitem[de Graauw et al.\ 1998]{de-Graauw-et-al-1998} de Graauw T., et al.\ 1998, SPIE, 3357, 336.
\bibitem[de Graauw et al.\ 2005]{de-Graauw-et-al-2005} de Graauw T., et al.\ 2005, \baas, 37, 1219.
\bibitem[Green \& Thaddeus 1974]{Green-and-Thaddeus-1974} Green S.\ \& Thaddeus P., 1974, ApJ, 191, 653.
\bibitem[Heyer \& Brunt 2004]{Heyer-and-Brunt-2004} Heyer M.\ H.\ \& Brunt C.\ M.\ 2004, ApJ, 615, L45.
\bibitem[Huttemeister-et-al-1995]{Huttemeister-et-al-1995} H\"{u}ttemeister S., Henkel C., Mauersberger R., Brouillet N., Wiklind T., Millar T.\ J.\ 1995, A\&A, 295, 571. 
\bibitem[Iwasawa et al. (2005)]{Iwasawa-et-al-2005} Iwasawa K., Sanders D.\ B., Evans A.\ S., Trentham N., Miniutti G., Spoon H.\ W.\ W.\ 2005, MNRAS, 357, 565.
\bibitem[Jackson et al.\ 1995]{Jackson-et-al-1995} Jackson J.\ M., Paglione T.\ A.\ D., Carlstrom J.\ E.\ \& Nguyen-Q R.\ 1995, ApJ, 438, 695.
\bibitem[Jansen-1995]{Jansen-1995} Jansen D.\ 1995, Ph.D.\ thesis, Sterrewacht, Universiteit Leiden (1995).
\bibitem[Komossa et al.\ (2003)]{Komossa-et-al-2003} Komossa S., Burwitz V., Hasinger G., Predehl P., Kaastra J.\ S., Ikebe Y.\ 2003, ApJ, 582, L15.
\bibitem[Krumholz \& McKee 2005]{Krumhol-and-McKee-2005} Krumholz M.\ R.\ \& McKee C.\ F.\ 2005, ApJ, 630, 250.
\bibitem[Kutner \& Ulich 1981]{Kutner-and-Ulich-1981} Kutner M.\ L.\ \& Ulich B.\ L.\ 1981\ ApJ, 250, 341.
\bibitem[Lahuis \& van Dishoeck 2000]{Lahuis-and-van-Dischoeck-2000} Lahuis F.\ \& van Dishoeck E.\ F.\ 2000, A\&A, 355, L699.
\bibitem[Langer and Penzias 1993]{Langer-and-Penzias-1993} Langer W.\ D.\ \& Penzias A.\ A.\ 1993, ApJ, 408, 539.
\bibitem[Larson 1981]{Larson-1981} Larson R.\ B.\ 1981, MNRAS, 194, 809.
\bibitem[Lepp \& Dalgarno 1996]{Lepp-and-Dalgarno-1996} Lepp S.\ \& Dalgarno A.\ 1996, A\&A, 306, L21.
\bibitem[Lisenfeld et al.\ 1996]{Lisenfeld-et-al-1996} Lisenfeld U., Hills R.\ E., Radford S.\ J.\ E., Solomon P.\ M.\ 1996, in 'Cold Gas at High Redshifts', p55, Kluwer Academic Publishers.
\bibitem[Lonsdale, Farrah \& Smith 2006]{Lonsdale-Farrah-Smith-2006} Lonsdale C., Farrah D.\ \& Smith H.\ 2006, 
in ``Astrophysics Update 2 - topical and timely reviews on astronomy and astrophysics". Ed. John W. Mason. 
Springer/Praxis books.
\bibitem[Mauersberger et al.\ 1999]{Mauersberger-et-al-1999} Mauersberger R., Henkel C., Walsh W., Schulz A.\ 1999, A\&A, 341, 256.
\bibitem[Nakanishi et al.\ 2005]{Nakanishi-et-al-2005} Nakanishi K., Okumura S.\ K., Kohno K., Kawabe R., Nakagaw T.\ 2005, PASJ, 4.
\bibitem[Nguyen et al.\ 1992]{Ngyyen-et-al-1992} Nguyen Q.-Rieu, Jackson J.M., Henkel C., Truong B., Mauersberger R.\ 1992, ApJ, 399, 521.
\bibitem[Paglione et al.\ 1997]{Paglione-et-al-1997} Paglione T.\ A.\ D., Jackson J.\ M., Ishizuki S.\ 1997, ApJ, 484, 656.
\bibitem[Papadopoulos, Seaquist \& Scoville (1996)]{Papadopoulos-Seaquist-Scoville-1996} Papadopoulos P.\ P., Seaquist E.\ R.\ \& Scoville N.\ Z.\ 1996, ApJ, 465, 173. 
\bibitem[Papadopoulos \& Seaquist 1998]{Papadopoulos-and-Seaquist-1998} Papadopoulos P.\ P.\ \& Seaquist E.\ R.\ 1998, ApJ, 492, 521. 
\bibitem[Papadopoulos \& Seaquist 1999]{Papadopoulos-and-Seaquist-1999} Papadopoulos P.\ P.\ \& Seaquist E.\ R.\ 1999, ApJ, 516, 114. 
\bibitem[Papadopoulos and Greve 2004]{Papadopoulos-and-Greve-2004} Papadopoulos P.\ P.\ \& Greve T.\ R.\ 2004, ApJ, 615, L29. 
\bibitem[Papadopoulos et al.\ 2007a]{Papadopoulos-et-al-2006} Papadopoulos P. P., Isaak K. G., \& van der Werf P. P. 2007a, ApJ, 668, 815.
\bibitem[Papadopoulos et a.\ 2007b] Papadopoulos P. P., Greve T. R., van der Werf P.  P.,  M\"uehle S. , Isaak, K., 
\&  Gao, Y.\ 2007b, Astrophysics \& Space Science, special issue for 2006 conference "Science with ALMA: a new era for Astrophysics" (arXiv:astro-ph/0701829).
\bibitem[Papadopoulos 2007]{Papadopoulos-2007} Papadopoulos P. P.\ 2007, ApJ, 656, 792
\bibitem[Plume et al.\ 1997]{Plume-et-al-1997} Plume R., Jaffe D.\ T., Evans N.\ J., Mart\'{i}n-Pintado J., G\'{o}mez-Gonz\'{a}lez J.\ 1997, ApJ, 476, 730.
\bibitem[Radford et al.\ 1991a]{Radford-et-al-1991a} Radford S.\ J.\ E., Solomon P.\ M.\ \& Downes D.\ 1991a, ApJ, 368, L15.
\bibitem[Radford et al.\ 1991b]{Radford-et-al-1991b} Radford S.\ J.\ E., et al.\ 1991b, IAUS, 146, 303.                 
\bibitem[Richardson 1985]{Richardson-1985} Richardson K.\ J.\ 1985, Ph.D.\ thesis, Dept.\ Phys., Queen Mary College, Univ. London, (1985).
\bibitem[Sakamoto et al. 2008]{Sakamoto-et-al-2008} Sakamoto K., et al.\ 2008, ApJ, in press. 
\bibitem[Sanders \& Mirabel 1996]{Sanders-and-Mirabel-1996} Sanders D.\ B.\ \& Mirabel I.\ F.\ 1996, ARA \& A 34, 749.
\bibitem[Schilke et al.\ 1992]{Schilke-et-al-1992} Schilke P., Walmsley C.\ M., Pineau Des Forets, G., Roueff E., Flower D.\ R., Guilloteau S.\ 1992, A\&A, 256, 595.
\bibitem[Scoville et al. 1991]{Scoville-et-al-1991} Scoville N.\ Z., Sargent A.\ I., Sanders D.\ B., Soifer B.\ T.\ 1991, ApJ, 366, L5.
\bibitem[Scoville et al.\ 1997]{Scoville-et-al-1997} Scoville N.\ Z., Yun M.\ S., Bryant P.\ M.\ 1997, ApJ, 484, 702.
\bibitem[Scoville 2003]{Scoville-2003} Scoville N.\ Z.\ 2003, proceedings of "The Neutral ISM in Starburst Galaxies", Marstrand, Sweden, Ed.\ S.\ Aalto, S.\ H\"{u}ttemeister \& A.\ Pedlar, p.\ 253-261. 
\bibitem[Seaquist, Lee \& Moriarty-Schieven 2006]{Seaquist-Lee-Moriarty-Schieven-2006} Seaquist E.\ R., Lee S.\ W., Moriarty-Schieven G.\ H.\ 2006, ApJ, 638, 148.
\bibitem[Shirley et al.\ 2003]{Shirley-et-al-2003} Shirley Y.\ L., Evans N.\ J., Young K.\ E., Knez C., Jaffe D.\ T.\ 2003, ApJS, 149, 375.
\bibitem[Soifer et al.\ 1986]{Soifer-et-al-1986} Soifer B.\ T., Sanders D.\ B., Neugebauer G., Danielson G.\ E., 
Lonsdale C.\ J., Madore B.\ F., Persson S.\ E.\ 1986, ApJ, 303, L41.
\bibitem[Solomon et al. 1990]{Solomon-et-al-1990} Solomon P.\ M., Radford S.\ J.\ E., Downes D.\ 1990, ApJ, 348, L53.
\bibitem[Solomon et al. 1992a]{Solomon-et-al-1992a} Solomon P.\ M., Downes D., Radford S.\ J.\ E.\ 1992a, ApJ, 387, L55.
\bibitem[Solomon et al. 1992b]{Solomon-et-al-1992b} Solomon P.\ M., Downes D., Radford S.\ J.\ E.\ 1992b, ApJ, 398, L29.
\bibitem[Solomon et al. 1997]{Solomon-et-al-1997} Solomon P.\ M., Downes D., Radford S.\ J.\ E., Barrett J.\ W., 1997, ApJ, 478, 144.
\bibitem[Spergel et al.\ 2003]{Spergel-et-al-2003} Spergel D.\ N., et al.\ 2003, ApJS, 148, 175.
\bibitem[Tacconi et al.\ 1999]{Tacconi-et-al-1999} Tacconi L.\ J., Genzel R., Tecza M., Gallimore J.\ F., Downes D., Scoville N.\ Z.\ 1999, ApJ, 524, 732.
\bibitem[Turner et al.\ 1992]{Turner-et-al-1992} Turner B.\ E., Chan K, Green S., Lubowich D.\ A., 1992, ApJ, 399, 114.
\bibitem[Usero et al. 2004]{Usero-et-al-2004} Usero A., Garc\'{i}a-Burillo S., Fuente A., Mart\'{i}n-Pintado J.\ 2004, ASPC. 320, 273.
\bibitem[Weiss-Walter-Scoville-2005]{Weiss-Walter-Scoville-2005} Wei\ss~A., Walter F., Scoville N.\ Z.\ 2005, A\&A, 438, 533.
\bibitem[Wiedner-et-al-2002]{Wiedner-et-al-2002} Wiedner M.\ C., Wilson C.\ D., Harrison A., Hills R.\ E., Lay O.\ P., Carlstrom J.\ E.\ 2002, ApJ, 581, 299.
\bibitem[Wolfire et al. 2003]{Wolfire-et-al-2003} Wolfire M.\ G., McKee C.\ F., Hollenbach D., Tielens A.\ G.\ G.\ M.\ 2003, ApJ, 587, 278. 
\bibitem[Wu et al.\ 2005]{Wu-et-al-2005} Wu J., Evans N.\ J., Gao Y., Solomon P.\ M., Shirley Y.\ L., Vanden Bout P.\ A.\ 2005, ApJ, 635, L173.
\end{thebibliography}
\end{document}